\documentclass[twocolumns]{aa}
\usepackage{graphicx}
\usepackage{mathrsfs}

\begin{document}

   \title{Quantifying the uncertainties of chemical evolution studies}

   \subtitle{I. The stellar initial mass function and the stellar lifetimes}

   \author{Donatella Romano,
          \inst{1}
          Cristina Chiappini,
	  \inst{2}
	  Francesca Matteucci,
	  \inst{3}
	  \and 
	  Monica Tosi
	  \inst{1}
          }

   \offprints{D. Romano}

   \institute{INAF\,--\,Osservatorio Astronomico di Bologna,
             Via Ranzani 1, I-40127 Bologna, Italy\\
             \email{donatella.romano@bo.astro.it, monica.tosi@bo.astro.it}
         \and
             INAF\,--\,Osservatorio Astronomico di Trieste,
             Via G.B. Tiepolo 11, I-34131 Trieste, Italy\\
	     \email{chiappini@ts.astro.it}
         \and
             Dipartimento di Astronomia, Universit\`a di Trieste,
             Via G.B. Tiepolo 11, I-34131 Trieste, Italy\\
             \email{matteucci@ts.astro.it}
             }

   \date{Received May 4, 2004; accepted September 9, 2004}

   \abstract{The stellar initial mass function and the stellar lifetimes are 
             basic ingredients of chemical evolution models, for which 
	     different recipes can be found in the literature. In this paper, 
	     we quantify the effects on chemical evolution studies of the 
	     uncertainties in these two parameters. We concentrate on chemical 
	     evolution models for the Milky Way, because of the large number 
	     of good observational constraints. Such chemical evolution models 
	     have already ruled out significant temporal variations for the 
	     stellar initial mass function in our own Galaxy, with the 
	     exception perhaps of the very early phases of its evolution. 
	     Therefore, here we assume a Galactic initial mass function 
	     constant in time. Through an accurate comparison of model 
	     predictions for the Milky Way with carefully selected data sets, 
	     it is shown that specific prescriptions for the initial mass 
	     function in particular mass ranges should be rejected. As far as 
	     the stellar lifetimes are concerned, the major differences among 
	     existing prescriptions are found in the range of very low-mass 
	     stars. Because of this, the model predictions widely differ for 
	     those elements which are produced mostly by very long-lived 
	     objects, as for instance $^3$He and $^7$Li. However, it is 
	     concluded that model predictions of several important observed 
	     quantities, constraining the plausible Galactic formation 
	     scenarios, are fairly robust with respect to changes in both the 
	     stellar mass spectrum and the stellar lifetimes. For instance, 
	     the metallicity distribution of low-mass stars is nearly 
	     unaffected by these changes, since its shape is dictated mostly 
	     by the time scale for thin-disk formation.

   \keywords{Galaxy: abundances -- Galaxy: evolution -- Galaxy: formation -- 
             stars: luminosity function, mass function -- stars: fundamental 
	     parameters
             }
   }

   \authorrunning{D. Romano et al.}

   \titlerunning{Quantifying the uncertainties of chemical evolution studies}

   \maketitle

%

   \section{Introduction}

   The formation and evolution of galaxies is one of the outstanding problems 
   of astrophysics. In the last decade, a great deal of observational work has 
   shed light on the production and distribution of chemical elements inside 
   the Galaxy (e.g. Edvardsson et al. 1993; Cayrel 1996; Nissen \& Schuster 
   1997; Gratton et al. 2000; Chen et al. 2003; Gratton et al. 2003; Ivans et 
   al. 2003; Reddy et al. 2003; Zoccali et al. 2003; Akerman et al. 2004), 
   often leading to an evolutionary scenario much more complicated than 
   assumed in many models. Even more recently, abundance data have accumulated 
   for external galaxies at both low and high redshift, thus providing 
   precious information on the chemical evolution of different types of 
   galaxies and on the early stages of galaxy evolution (e.g. Pettini 2001; 
   Centuri\'on et al. 2003; Dietrich et al. 2003a, b; Prochaska, Howk \& Wolfe 
   2003; Tolstoy et al. 2003; Dessauges-Zavadsky et al. 2004; D'Odorico et al. 
   2004). In this framework, galactic chemical evolution models can be 
   regarded as useful tools to discriminate among different scenarios of 
   galaxy formation. In fact, as stressed many times in the literature, 
   abundances and abundance ratios play a major r\^ole as cosmic clocks and 
   give hints on the time scales of structure formation and evolution 
   (Wheeler, Sneden \& Truran 1989; Matteucci \& Fran\c cois 1992; Matteucci 
   2001).

   In order to build up a chemical evolution model, it is necessary to define 
   the initial conditions and the basic physical laws governing the evolution 
   of the system during the whole galactic lifetime. In short, one needs to 
   specify whether the system is closed or open (should any inflow/outflow of 
   gas occur), the chemical composition of the gas from which the computation 
   starts and that of any infalling material, the stellar birthrate function 
   and the stellar evolution and nucleosynthesis. In particular, the stellar 
   birthrate is expressed as the product of two independent functions, the 
   star formation rate (SFR) and the stellar initial mass function (IMF). The 
   first is generally expressed as a function of time only, while the second, 
   which describes the stellar mass distribution at birth,  is likely to be 
   universal (Kroupa 2002; but see e.g. Jeffries et al. 2004) and not vary as 
   a function of time (Chiappini, Matteucci \& Padoan 2000).

   The free parameters that one introduces in the model simply reflect our 
   poor understanding of the basic physical processes governing the formation 
   and evolution of galactic structures. However, having a number of 
   observational constraints formally larger than that of the free parameters 
   allows us to restrict the range of variation of the parameters themselves 
   and gain useful insight into the mechanisms of galaxy formation and 
   evolution (Matteucci 2001). This is the case for our own Galaxy and will 
   become soon the standard for an increasing number of external galaxies, 
   thanks to the capabilities of modern telescopes and instrumentation.

   In an epoch where the uncertainties in the data have become really small, 
   it is worth trying to consider `theoretical error bars' as well. An attempt 
   to do so has recently been done by Romano et al. (2003), who compare the 
   evolution of light elements predicted by two independent models of chemical 
   evolution for the Milky Way and try to ascertain the origin of the 
   differences in the model predictions. In the present paper we intend to 
   assess the uncertainties in the model predictions which arise when 
   exploring different prescriptions for the IMF and the stellar lifetimes in 
   a model for the Milky Way. To this purpose, we adopt a chemical evolution 
   model which has been proven to successfully reproduce the main 
   observational features of the solar neighbourhood (Chiappini, Matteucci \& 
   Gratton 1997) and change the assumptions on the stellar IMF. As a result, 
   we set up a range of possible variations for several predicted quantities. 
   Then, we repeat the same kind of analysis by changing the prescriptions for 
   the stellar lifetimes.

   The paper is organized as follows. In Section~2, we review the general 
   features of the IMF which have emerged over the years and emphasize 
   differences and similarities among the various IMFs adopted in this work. 
   In Section~3, we discuss the prescriptions for the stellar lifetimes. In 
   Section~4, we describe the chemical evolution model for the solar vicinity. 
   In Section~5 we present model results. Finally, Section~6 is devoted to a 
   critical discussion of the problem and some conclusions are drawn. Notice 
   that the paper is structured in such a way that it is easy to concentrate 
   on only one among the proposed topics, while skipping the others, if one 
   wishes to do so.

   \section{The stellar initial mass function}

   \subsection{The adopted parametrizations}

%
   \begin{table*}
   \caption[]{${\mathrm I}(m_1,\, m_2)$ (see text for a definition of this 
              quantity) for different mass intervals for the various IMFs 
	      considered in this study.}
   \begin{center}
   \begin{tabular}{c c c c c c c}
   \hline \hline
   $\mathrm{Mass\; range}$ & 
   \multicolumn{6}{c}{${\mathrm I}(m_1,\, m_2)$} \\
   \cline{2-7}
   $(m_1,\, m_2)$ & ${\mathrm{Salpeter\; (1955)}}$ & ${\mathrm{Tinsley\; 
   (1980)}}$ & ${\mathrm{Scalo\; (1986)}}$  & ${\mathrm{Kroupa\; et\; al. 
   (1993)}}$ & ${\mathrm{Scalo\; (1998)}}$ & ${\mathrm{Chabrier\; (2003)}}$ \\
   \cline{1-7}
   $0.1$--$0.5$ & $0.472814$  & $0.340757$   & $0.52945$  & $0.343983$  & 
   $0.205694$  & $0.246036$  \\
   $0.5$--$0.6$ & $0.0386389$ & $0.0386018$  & $0.0432673$ & $0.0637846$ & 
   $0.0446104$  & $0.0404886$ \\
   $0.6$--$1.0$ & $0.0960027$ & $0.108154$    & $0.107502$  & $0.166806$  & 
   $0.165927$  & $0.119639$  \\
   $1.0$--$2.0$ & $0.105638$  & $0.146756$   & $0.118292$  & $0.170327$  & 
   $0.217311$  & $0.148891$  \\
   $2.0$--$5.0$ & $0.105561$  & $0.169621$   & $0.10199$  & $0.129128$  & 
   $0.164747$  & $0.154816$  \\
   $5.0$--$8.0$ & $0.0423486$ & $0.0705069$  & $0.0318011$ & $0.0402628$ & 
   $0.0513691$ & $0.0643529$ \\
   $8.0$--$40$  & $0.102002$  & $0.184335$   & $0.0551705$ & $0.0698502$ & 
   $0.108612$  & $0.162753$  \\
   $40$--$100$  & $0.0369948$ & $0.00322992$ & $0.0125267$ & $0.0158598$ & 
   $0.0417308$ & $0.0630242$ \\
   \cline{1-7}
   \end{tabular}
   \end{center}
   \end{table*}
%

%
   \begin{figure*}
   \centering
   \includegraphics[width=\columnwidth]{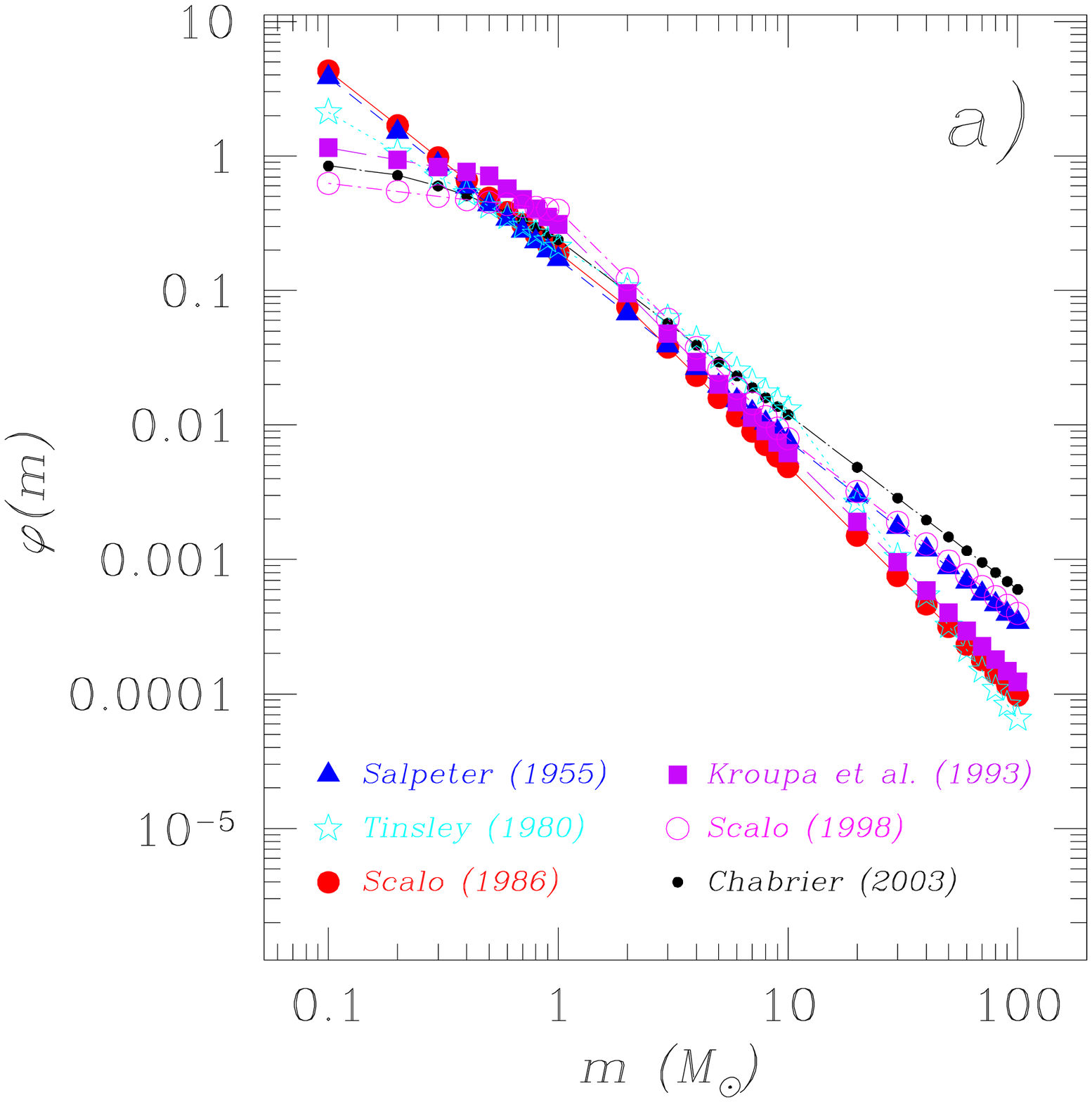}
   \includegraphics[width=\columnwidth]{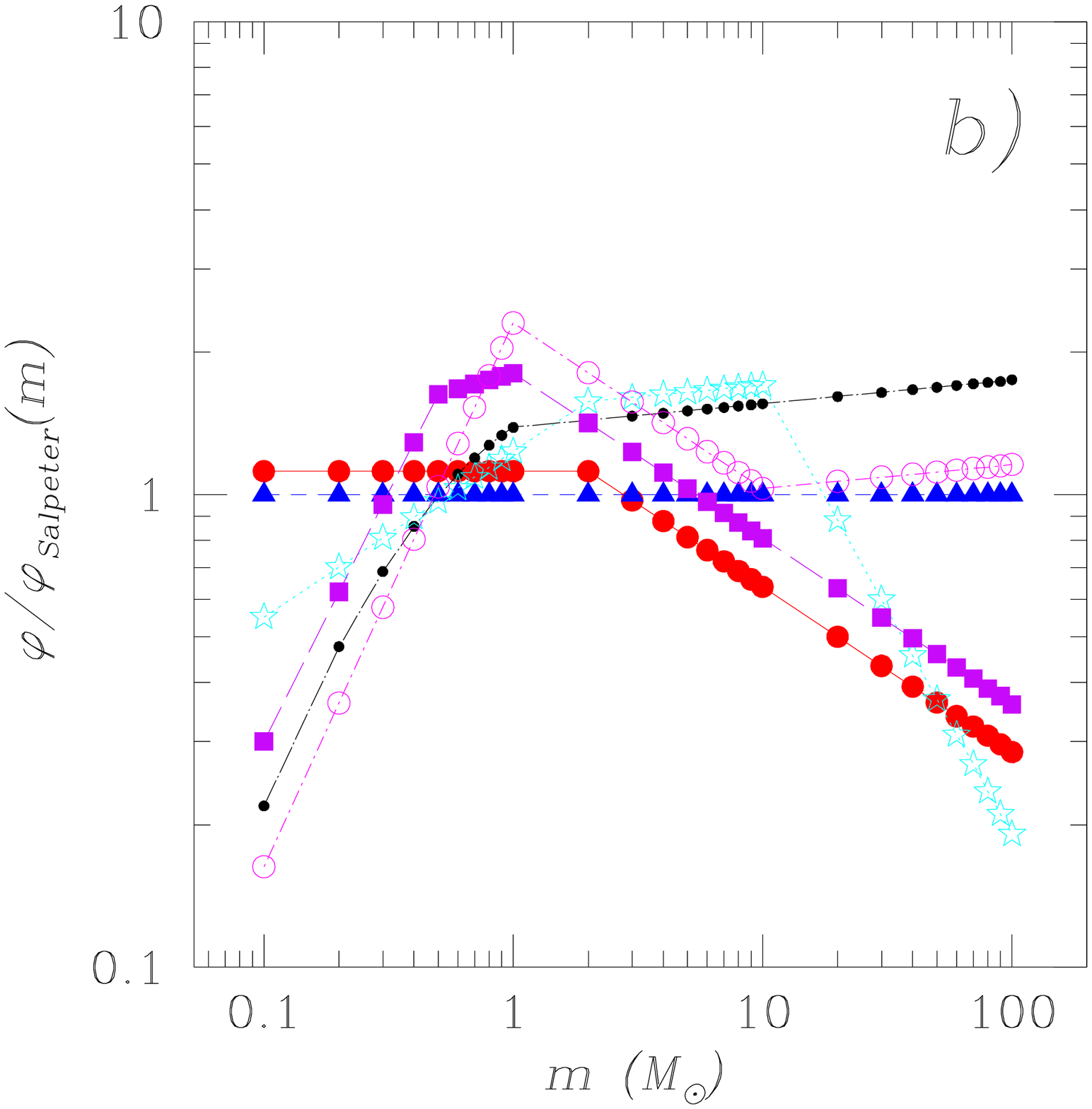}
      \caption{a) Stellar IMF according to Salpeter (1955; {\it triangles}), 
	       Tinsley (1980; {\it stars}), Scalo (1986; {\it full circles}), 
	       Kroupa et al. (1993; {\it squares}), Scalo (1998; {\it empty 
	       circles}) and Chabrier (2003; {\it dots}). Here $\varphi(m)$ is 
	       the IMF by mass and $\varphi(m) \propto m^{-x}$ ($x$ = 1.35 in 
	       the case of Salpeter's IMF), except for Chabrier (2003), where 
	       a lognormal form for the low-mass domain ($m \le$ 1 $M_\odot$) 
	       is suggested instead (see text). Here the Chabrier IMF also has 
	       $x$ = 1.3 as the exponent in the $m >$ 1 $M_\odot$ mass domain. 
	       b) Same as panel a), but with $\varphi(m)$ divided by the 
	       corresponding Salpeter value for each given mass. This allows a 
	       first sight comparison of the various mass distributions 
	       expected according to the different IMFs with respect to the 
	       `standard' Salpeter choice.}
         \label{FigIMF1}
   \end{figure*}
%

   The most widely used functional form for the IMF is an extension of that 
   proposed by Salpeter (1955) to the whole stellar mass range:
   \begin{displaymath}
     \phi_{\mathrm{Salpeter}}(m) = \mathscr{A}_{\mathrm{Salpeter}}\, 
                                   m^{-(1 + x)},
   \end{displaymath}
   where $x$ = 1.35 and
   \begin{displaymath}
     \int_{m_{\mathrm{low}}}^{m_{\mathrm{up}}} m\,\phi_{\mathrm{Salpeter}}(m) 
     {\mathrm{d}}m = 1.
   \end{displaymath}
   Here the IMF is by number; $m_{\mathrm{low}}$ = 0.1 $M_\odot$, 
   $m_{\mathrm{up}}$ = 100 $M_\odot$ and $\mathscr{A}_{\mathrm{Salpeter}} 
   \simeq$ 0.17. The above formula equivalently reads
   \begin{displaymath}
     \int_{m_{\mathrm{low}}}^{m_{\mathrm{up}}} \varphi_{\mathrm{Salpeter}}(m) 
     {\mathrm{d}}m = 1
   \end{displaymath}
   if using the IMF by mass, $\varphi_{\mathrm{Salpeter}}(m) \propto 
   m^{-1.35}$. In what follows we always use the IMF by mass.

   More realistic, multi-slope expressions better describe the luminosity 
   function of main sequence stars in the solar neighbourhood, that is what is 
   actually observed (Tinsley 1980; Scalo 1986; Kroupa, Tout \& Gilmore 1993; 
   Scalo 1998; see the original publications for details):
   \begin{displaymath}
     \varphi_{\mathrm{Tinsley}}(m) = \left\{ \begin{array}{l l}
                             \mathscr{A}_{\mathrm{Tinsley}}\, m^{-1.0} & 
			     \qquad {\mathrm{if}} \; m < 2 \, M_\odot \\
                             \mathscr{B}_{\mathrm{Tinsley}}\, m^{-1.3} & 
			     \qquad {\mathrm{if}} \; 2 < m/M_\odot < 10 \\
                             \mathscr{C}_{\mathrm{Tinsley}}\, m^{-2.3} & 
			     \qquad {\mathrm{if}} \; m > 10 \, M_\odot, \\
                                     \end{array} \right.
   \end{displaymath}
   \begin{displaymath}
   \mathscr{A}_{\mathrm{Tinsley}} \simeq 0.21, \;
   \mathscr{B}_{\mathrm{Tinsley}} \simeq 0.26, \;
   \mathscr{C}_{\mathrm{Tinsley}} \simeq 2.6;
   \end{displaymath}
   \begin{displaymath}
     \varphi_{\mathrm{Scalo\, 86}}(m) = \left\{ \begin{array}{l l}
                             \mathscr{A}_{\mathrm{Scalo\, 86}}\, m^{-1.35} & 
			     \qquad {\mathrm{if}} \; m < 2 \, M_\odot \\
                             \mathscr{B}_{\mathrm{Scalo\, 86}}\, m^{-1.70} & 
			     \qquad {\mathrm{if}} \; m > 2 \, M_\odot, \\
                                        \end{array} \right.
   \end{displaymath}
   \begin{displaymath}
   \mathscr{A}_{\mathrm{Scalo\, 86}} \simeq 0.19, \;
   \mathscr{B}_{\mathrm{Scalo\, 86}} \simeq 0.24
   \end{displaymath}
   (notice that we adopt a simplified two-slope approximation to the actual 
   Scalo 1986 formula, similarly to what is done in Matteucci \& Fran\c cois 
   1989);
   \begin{displaymath}
     \varphi_{\mathrm{Kroupa}}(m) = \left\{ \begin{array}{l l}
                             \mathscr{A}_{\mathrm{Kroupa}}\, m^{-0.3} & \qquad 
				     {\mathrm{if}} \; m < 0.5 \, M_\odot \\
                             \mathscr{B}_{\mathrm{Kroupa}}\, m^{-1.2} & \qquad 
				     {\mathrm{if}} \; 0.5 < m/M_\odot < 1 \\
                             \mathscr{C}_{\mathrm{Kroupa}}\, m^{-1.7} & \qquad 
				     {\mathrm{if}} \; m > 1 \, M_\odot, \\
                                     \end{array} \right.
   \end{displaymath}
   \begin{displaymath}
   \mathscr{A}_{\mathrm{Kroupa}} \simeq 0.58, \;
   \mathscr{B}_{\mathrm{Kroupa}} = \mathscr{C}_{\mathrm{Kroupa}} \simeq 0.31;
   \end{displaymath}
   \begin{displaymath}
     \varphi_{\mathrm{Scalo\, 98}}(m) = \left\{ \begin{array}{l l}
                             \mathscr{A}_{\mathrm{Scalo\, 98}}\, m^{-0.2} & 
			     \qquad {\mathrm{if}} \; m < 1 \, M_\odot \\
                             \mathscr{B}_{\mathrm{Scalo\, 98}}\, m^{-1.7} & 
			     \qquad {\mathrm{if}} \; 1 < m/M_\odot < 10 \\
                             \mathscr{C}_{\mathrm{Scalo\, 98}}\, m^{-1.3} & 
			     \qquad {\mathrm{if}} \; m > 10 \, M_\odot, \\
                                        \end{array} \right.
   \end{displaymath}
   $\mathscr{A}_{\mathrm{Scalo\, 98}} = \mathscr{B}_{\mathrm{Scalo\, 98}} 
   \simeq$ 0.39, $\mathscr{C}_{\mathrm{Scalo\, 98}} \simeq$ 0.16. All of them 
   are considered in the present work. The normalization is always performed 
   in the mass range 0.1--100 $M_\odot$.

   More recently, a lognormal form has been suggested for the low-mass part of 
   the IMF ($m \le$ 1 $M_\odot$), eventually extending into the substellar 
   regime (Chabrier 2003):
   \begin{displaymath}
     \varphi_{\mathrm{Chabrier}}(m) = \left\{ \begin{array}{l}
                             \mathscr{A}_{\mathrm{Chabrier}}\, 
				     {\mathrm{e}}^{-(\log m - \log 
				     m_{\mathrm{c}})^2/2\,\sigma^2} \\
				     \qquad \qquad \qquad \qquad \qquad 
					    {\mathrm{if}} \;m \le 1 \, M_\odot 
					    \\
                             \mathscr{B}_{\mathrm{Chabrier}}\, 
			             m^{-(1.3 \pm 0.3)} \\
				     \qquad \qquad \qquad \qquad \qquad 
					    {\mathrm{if}} \; m > 1 \, M_\odot. 
					    \\
                                      \end{array} \right.
   \end{displaymath}
   According to Chabrier (2003), the IMF depends weakly on the environment, 
   except perhaps for early star formation conditions. Values of 
   $m_{\mathrm{c}}$ = 0.079 $M_\odot$ and $\sigma$ = 0.69 well characterize 
   the IMF for single objects belonging to the Milky Way disk. For $m > 1$ 
   $M_\odot$, we assume a power-law exponent $x$ equal to either 1.3 or 1.7. 
   In the latter case, we study both an IMF truncated at $m_{\mathrm{low}}$ = 
   0.1 $M_\odot$ and one extending down to $m_{\mathrm{low}}$ = 0.001 
   $M_\odot$, i.e., into the brown dwarf (BD) domain. The normalization 
   constants are: $\mathscr{A}_{\mathrm{Chabrier}} \simeq$ 0.85, 
   $\mathscr{B}_{\mathrm{Chabrier}} \simeq$ 0.24 for $x$ = 1.3, 
   $m_{\mathrm{low}}$ = 0.1 $M_\odot$; $\mathscr{A}_{\mathrm{Chabrier}} 
   \simeq$ 1.16, $\mathscr{B}_{\mathrm{Chabrier}} \simeq$ 0.32 for $x$ = 1.7, 
   $m_{\mathrm{low}}$ = 0.1 $M_\odot$; $\mathscr{A}_{\mathrm{Chabrier}} 
   \simeq$ 1.06, $\mathscr{B}_{\mathrm{Chabrier}} \simeq$ 0.30 for $x$ = 1.7, 
   $m_{\mathrm{low}}$ = 0.001 $M_\odot$.

%
   \begin{table*}
   \caption[]{${\mathrm I}(m_1,\, m_2)$ for different mass ranges in case of 
              Chabrier (2003) IMF with $x$ = 1.7. The normalization is 
	      performed either in the mass range 0.1--100 $M_\odot$ (second 
	      column) or in the mass range 0.001--100 $M_\odot$ (third 
	      column).}
   \begin{center}
   \begin{tabular}{c c c}
   \hline \hline
   $\mathrm{Mass\; range}$ & 
   \multicolumn{2}{c}{${\mathrm I}(m_1,\, m_2)$} \\
   \cline{2-3}
   $(m_1,\, m_2)$ & ${\mathrm{Chabrier\; (2003);\; steeper,\; without\; BDs}}$ 
   & ${\mathrm{Chabrier\; (2003);\; steeper,\; with\; BDs}}$ \\
   \cline{1-3}
   $0.001$--$0.1$ & $-$         & $0.0882928$ \\
   $0.1$--$0.5$   & $0.335882$  & $0.306225$  \\
   $0.5$--$0.6$   & $0.0552738$ & $0.0503936$ \\
   $0.6$--$1.0$   & $0.163328$  & $0.148907$  \\
   $1.0$--$2.0$   & $0.178369$  & $0.162621$  \\
   $2.0$--$5.0$   & $0.135225$  & $0.123286$  \\
   $5.0$--$8.0$   & $0.0421639$ & $0.0384412$ \\
   $8.0$--$40$    & $0.0731485$ & $0.0666901$ \\
   $40$--$100$    & $0.0166087$ & $0.0151423$ \\
   \cline{1-3}
   \end{tabular}
   \end{center}
   \end{table*}
%

%
   \begin{figure}
   \centering
   \includegraphics[width=\columnwidth]{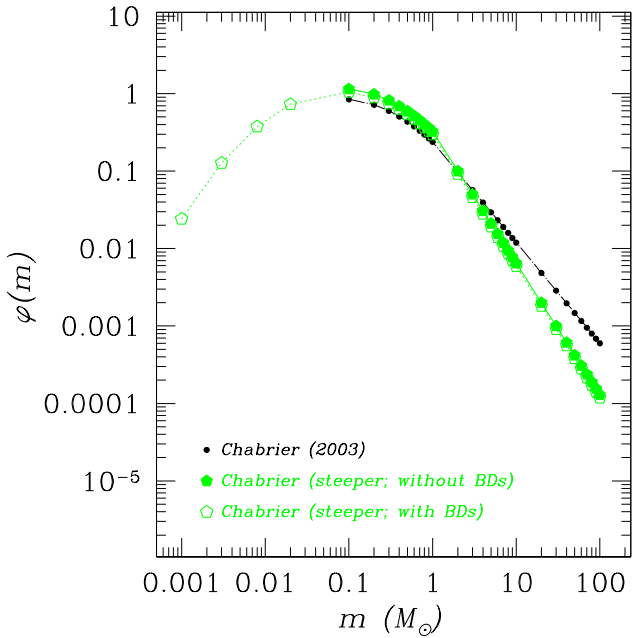}
      \caption{Stellar IMF according to Chabrier (2003). A power law is 
               assumed for $m \ge$ 1 $M_\odot$, with an exponent $x$ = 1.3 
	       {\it (dots)} or $x$ = 1.7 {\it (pentagons)}. For $m <$ 1 
	       $M_\odot$, a lognormal form is proposed, eventually extending 
	       into the BD domain {\it (open pentagons)}.
               }
         \label{FigIMF2}
   \end{figure}
%

   The question of whether the IMF is universal or varies with varying 
   star-forming conditions has been long debated in the past. Combining recent 
   IMF estimates for different populations in which individual stars have been 
   resolved unveils an extraordinary uniformity of the IMF (Kroupa 2002), 
   although some room for exceptions is left (e.g., Aloisi, Tosi \& Greggio 
   1999). Explaining the chemo-photometric properties of elliptical galaxies 
   has often required an IMF biased towards massive stars (e.g. Arimoto \& 
   Yoshii 1987). However, the actual IMF slope cannot be inferred from direct 
   observations in this case.

   \subsection{Differences and similarities among different parametrizations}

   In Fig.~\ref{FigIMF1}a we compare Salpeter (1955; {\it triangles}), Tinsley 
   (1980; {\it stars}), Scalo (1986; {\it full circles}), Kroupa et al. (1993; 
   {\it squares}), Scalo (1998; {\it empty circles}) and Chabrier (2003; {\it 
   dots}) IMFs. For the Chabrier IMF the dots for $m \ge$ 1 $M_\odot$ display 
   the power-law form with an exponent $x =$ 1.3. For all these IMFs, the 
   normalization is performed in the mass range 0.1--100 $M_\odot$.

   It is immediately seen that a Salpeter or a Scalo (1986) IMF predicts many 
   more stars at the very low end of the distribution than a Scalo (1998) or a 
   Chabrier IMF. Tinsley's IMF lies somewhere in the middle, while it predicts 
   the highest fraction of stars in the mass range 2--10 $M_\odot$. These 
   features appear more clearly in Fig.~\ref{FigIMF1}b, where the IMFs are 
   divided by the corresponding Salpeter value for each given mass.

   In order to quantify the relative weights of stars belonging to different 
   mass ranges according to different IMFs, we compute the following 
   quantities:
   \begin{displaymath}
     {\mathrm{I}}(m_1, m_2) = \int_{m_1}^{m_2} \varphi(m) {\mathrm{d}}m,
   \end{displaymath}
   for each of the IMFs discussed above. The results are listed in Table~1. 
   The integration limits, $m_1$ and $m_2$, are properly chosen in order to 
   allow a meaningful comparison when discussing the r\^ole played by stars 
   belonging to different mass intervals in polluting the interstellar medium 
   (ISM) with different chemical elements while varying the IMF prescriptions 
   (Section~5.1).

   In Fig.~\ref{FigIMF2} we show the effect of steepening Chabrier (2003) IMF 
   for $m \ge$ 1 $M_\odot$. A value of $x$ = 1.7 {\it (pentagons)} is now 
   assumed instead of $x$ = 1.3 {\it (dots)}. A steeper IMF for stars more 
   massive than a few solar masses seems to be more likely, on the grounds of 
   recent results by Kroupa \& Weidner (2003). Those authors argue that field 
   IMFs for early-type stars must be steeper than the Salpeter approximation, 
   owing to the fact that a Salpeter power law well describes the distribution 
   of stellar masses for local star formation events in star clusters. A steep 
   field-star IMF thus arises naturally because of the power-law cluster mass 
   function according to which star clusters distribute.

%
   \begin{figure*}
   \centering
   \includegraphics[width=\columnwidth]{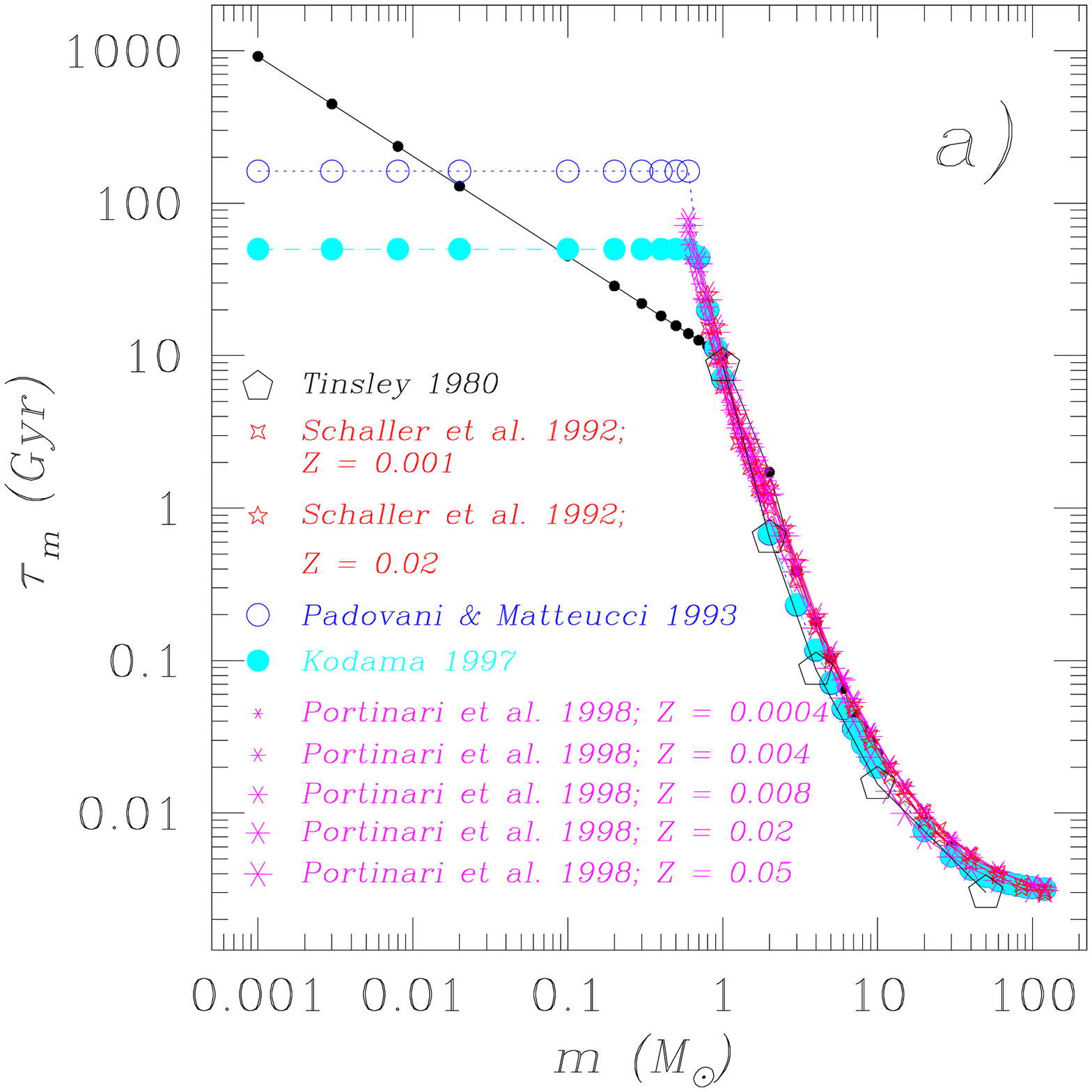}
   \includegraphics[width=\columnwidth]{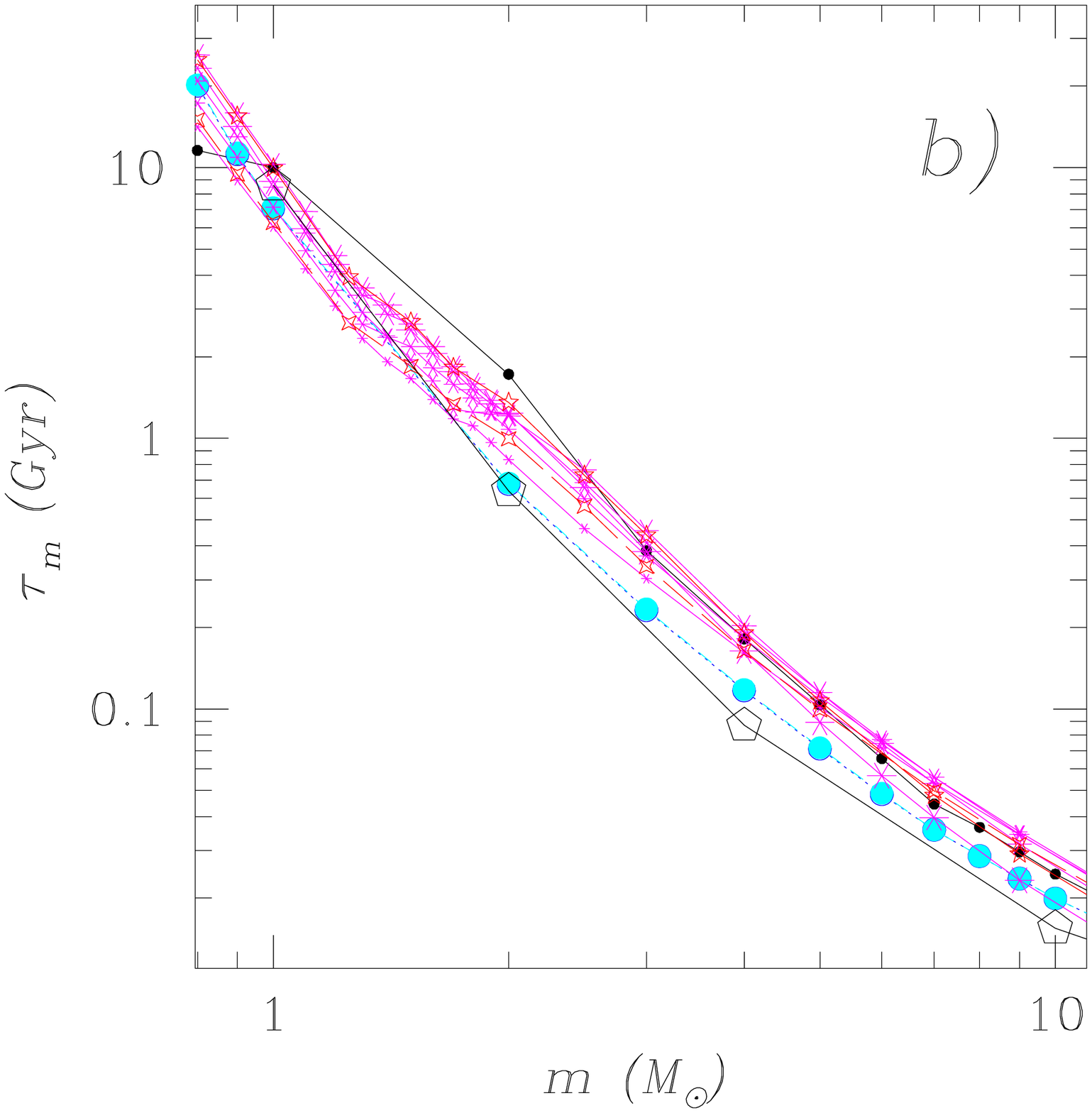}
      \caption{a) Stellar lifetimes as a function of the initial mass of the 
	       star: Maeder \& Meynet (1989; {\it dots}); Tinsley (1980; {\it 
	       pentagons}); Schaller et al. (1992; {\it stars}); Padovani \& 
	       Matteucci (1993; {\it empty circles}); Kodama (1997; {\it 
	       filled circles}); Portinari et al. (1998; {\it asterisks}). 
	       b) A zoom of the 1--10 $M_\odot$ stellar mass range.
               }
         \label{FigTAU}
   \end{figure*}
%

   Fig.~\ref{FigIMF2} also shows the behaviour of Chabrier IMF for $m <$ 1 
   $M_\odot$. It is seen that the mass distribution flattens below $m \simeq$ 
   1 $M_\odot$, reaches a peak around $m \simeq$ 0.1 $M_\odot$ and then 
   decreases smoothly for $m <$ 0.1 $M_\odot$. As a consequence, less than 
   10\% of the stellar mass should be found in form of BDs (see also Table~2).

   In Section~5.1 we illustrate the results obtained for the solar vicinity 
   when the different parametrizations listed above are adopted.

   \section{The stellar lifetimes}

   Different authors traditionally adopt different prescriptions for the 
   stellar lifetimes. For instance, Matteucci and coworkers have usually 
   assumed stellar lifetimes from Maeder \& Meynet (1989):
   \begin{displaymath}
     \tau_m = \left\{ \begin{array}{l l}
     10^{-0.6545 \log m + 1}  & \quad {\mathrm{for\;}} m \le 1.3\; M_\odot \\
     10^{-3.7 \log m + 1.35}  & \quad {\mathrm{for\;}} 1.3 < m/M_\odot \le 3 \\
     10^{-2.51 \log m + 0.77} & \quad {\mathrm{for\;}} 3 < m/M_\odot \le 7 \\
     10^{-1.78 \log m + 0.17} & \quad {\mathrm{for\;}} 7 < m/M_\odot \le 15 \\
     10^{-0.86 \log m - 0.94} & \quad {\mathrm{for\;}} 15 < m/M_\odot \le 60 \\
     1.2 \, m^{-1.85} + 0.003 & \quad {\mathrm{for\;}} m > 60\; M_\odot, \\
              \end{array} \right.
   \end{displaymath}
   with $\tau_m$ in units of Gyr. Notice that they extrapolated the values for 
   $m \le$ 1.3 $M_\odot$ and $m >$ 60 $M_\odot$, since Maeder \& Meynet (1989) 
   do not give any formula for these mass ranges. The adopted extrapolation to 
   larger masses is compatible with the values reported by Maeder \& Meynet 
   for stars of 85 and 120 $M_\odot$ (their table 2).

   These stellar lifetimes differ widely from other formulations available in 
   the literature, especially in the low-mass range ($m <$ 1 $M_\odot$). 
   Tinsley (1980) and Tosi (1982 and subsequent papers) have:
   \begin{displaymath}
     \begin{array}{l l}
     \tau_m > 8.6 & \quad {\mathrm{for\;}} m < 1\; M_\odot \\
     8.6 > \tau_m > 0.64 & \quad {\mathrm{for\;}} 1 < m/M_\odot < 2 \\
     0.64 > \tau_m > 0.087 & \quad {\mathrm{for\;}} 2 < m/M_\odot < 4 \\
     0.087 > \tau_m > \sim 0.0155 & \quad {\mathrm{for\;}} 4 < m/M_\odot < 10 
                                                                            \\
     \sim 0.0155 > \tau_m > \sim 0.003 & \quad {\mathrm{for\;}} 10 < m/M_\odot 
                                                                        < 50 \\
     \end{array}
   \end{displaymath}
   (see Tinsley 1980 for references).

   Kodama (1997) reports:
   \begin{displaymath}
     \tau_m = \left\{ \begin{array}{l}
     50 \quad {\mathrm{for\;}} m \le 0.56\; M_\odot \\
     10^{\big(0.334 - \sqrt{1.790 - 0.2232 \times (7.764 - 
     \log m)}\big)\big/0.1116} \\ 
     \qquad \qquad \qquad \qquad \qquad \qquad \qquad {\mathrm{for\;}} m \le 
     6.6\; M_\odot \\
     1.2 \, m^{-1.85} + 0.003 \quad {\mathrm{for\;}} m > 6.6\; M_\odot, \\
              \end{array} \right.
   \end{displaymath}
   similarly to what suggested by Padovani \& Matteucci (1993 and references 
   therein):
   \begin{displaymath}
     \tau_m = \left\{ \begin{array}{l}
     160 \quad {\mathrm{for\;}} m \le 0.60\; M_\odot \\
     10^{\big(0.334 - \sqrt{1.790 - 0.2232 \times (7.764 - \log 
     m)}\big)\big/0.1116} \\ 
     \qquad \qquad \qquad \qquad \qquad \qquad \qquad {\mathrm{for\;}} m \le 
     6.6\; M_\odot \\
     1.2 \, m^{-1.85} + 0.003 \quad {\mathrm{for\;}} m > 6.6\; M_\odot. \\
   \end{array} \right.
   \end{displaymath}
   It is seen that ancient very low-mass stars (0.6 $\le m/M_\odot \le$ 0.9) 
   die in a Hubble time according to Matteucci \& Fran\c cois (1989) 
   extrapolation of Maeder \& Meynet (1989) formul\ae, whereas they have a 
   longer life according to other authors. On the contrary, following Maeder 
   \& Meynet (1989) stars with 1 $< m/M_\odot <$ 2 live longer than in any 
   other study.

   In Fig.~\ref{FigTAU} we compare all the above-mentioned parametrizations, 
   as well as values obtained from Geneva (Schaller et al. 1992) and Padua 
   (Portinari et al. 1998 and references therein) stellar tracks. In the low- 
   and very low-mass stellar mass domain substantial differences are found, 
   while smaller, though not negligible, differences exist at higher masses. 
   Detailed analyses of the results obtained when integrating these 
   differences over the Galactic lifetime and IMF are reported in Section~5.2.

   \section{The chemical evolution model for the solar neighbourhood}

   In order to follow the chemical evolution of the solar vicinity, we adopt 
   the two-infall model of Chiappini et al. (1997). According to this model, 
   the halo and part of the thick-disk population form out of a first infall 
   episode on a short time scale, while the thin disk accumulates much more 
   slowly during a second independent infall episode. The rate of accretion of 
   matter is
   \begin{displaymath}
     \frac{\mathrm{d}\Sigma_{\mathrm{inf}}(t)}{\mathrm{d}t} = A \, 
     \mathrm{e}^{-t/\tau_{\mathrm{H}}} + B \, \mathrm{e}^{-(t - 
     t_{\mathrm{max}})/\tau_{\mathrm{D}}},
   \end{displaymath}
   where $\Sigma_{\mathrm{inf}}(t)$ is the mass surface density of the 
   infalling primordial matter at time $t$. The parameters $t_{\mathrm{max}}$ 
   = 1 Gyr, $\tau_{\mathrm{H}}$ = 0.8 Gyr and $\tau_{\mathrm{D}}$ = 7 Gyr are 
   the time of maximum infall onto the disk and the time scales for mass 
   accretion onto the halo/thick-disk and thin-disk components, respectively. 
   They are fixed by the request of reproducing a number of observational 
   constraints (Matteucci 2001). The coefficients $A$ and $B$ are derived from 
   the condition of reproducing the current total mass surface density at the 
   solar position (Chiosi 1980). Obviously, the coefficient $B$ must be zero 
   for $t < t_{\mathrm{max}}$.

   The SFR is
   \begin{displaymath}
     \psi(t) = \nu(t) \bigg[ \frac{\Sigma(t_{\mathrm{Gal}})}{\Sigma(t)} 
               \bigg]^{k-1} G^k(t),
   \end{displaymath}
   proportional to both the total mass and gas surface densities. Here $G(t)$ 
   is the normalized gas surface density, 
   $\Sigma_{\mathrm{gas}}(t)/\Sigma(t_{\mathrm{Gal}})$, and 
   $\Sigma(t_{\mathrm{Gal}})$ is the total mass surface density at the present 
   time, $t_{\mathrm{Gal}}$ = 13.7 Gyr\footnote{Notice that here we adopt a 
   younger Galaxy age to allow for consistency with the recent \emph{WMAP} 
   data on the universe age (see also Romano et al. 2003).}. A gas exponent 
   $k$ = 1.5 guarantees a good agreement between model predictions and 
   observations (Chiappini et al. 1997). Moreover, it is found to agree with 
   inferences from observations (Kennicutt 1998) and $N$-body simulations 
   (Gerritsen \& Icke 1997).

   The star formation efficiency, $\nu(t)$, is set to 2 Gyr$^{-1}$ during the 
   halo/thick-disk phase and to 1.2 Gyr$^{-1}$ during the thin-disk phase to 
   ensure the best fit to all the observed features of the solar vicinity, 
   unless the gas surface density drops below a critical threshold, 
   $\Sigma_{\mathrm{gas}}^{\mathrm{th}} \sim$ 7 $M_\odot$ pc$^{-2}$. In this 
   case $\nu$ = 0 and the star formation ceases. This naturally explains the 
   existence of a gap in the SFR between the halo/thick-disk and the thin-disk 
   phase (Gratton et al. 1996, 2000; Fuhrmann 1998, 2004). Moreover, it delays 
   the beginning of the star formation in the halo to the time at which the 
   critical gas density can be reached.

   The instantaneous recycling approximation is relaxed, i.e. the stellar 
   lifetimes are taken into account in details. Stellar nucleosynthesis is 
   taken from (i) van den Hoek \& Groenewegen (1997) for low- and 
   intermediate-mass stars (their case with metallicity-dependent mass loss 
   efficiency along the asymptotic giant branch); (ii) Woosley \& Weaver 
   (1995) for massive stars; (iii) Thielemann, Nomoto \& Hashimoto (1993) for 
   Type Ia supernovae (SNeIa); (iv) Jos\'e \& Hernanz (1998) for classical 
   novae. The carbon yields from massive stars in the mass range $m \ge$ 40 
   $M_\odot$ are multiplied by a factor of three (arguments are given in 
   Chiappini, Romano \& Matteucci 2003a). Stellar production and/or 
   destruction of the light elements deuterium, $^3$He and $^4$He are included 
   in the model according to Dearborn, Steigman \& Tosi (1996), Galli et al. 
   (1997) and Chiappini, Renda \& Matteucci (2002) prescriptions. $^7$Li 
   production/destruction is accounted for following Romano et al. (2001, 
   2003) and references therein.

   The prescriptions on the IMF and the stellar lifetimes are changed 
   according to the above discussions (Sections~2 and 3).

   \section{Quantifying the uncertainties in chemical evolution model 
            predictions} 

   \subsection{Changing the IMF}  

   In this section we discuss the results obtained by varying the 
   prescriptions on the stellar IMF in the chemical evolution code for the 
   solar vicinity. The purpose is to associate errors due to uncertainties in 
   the stellar IMF to chemical evolution model results. 

   During the Galactic lifetime, many successive stellar generations form 
   according to the adopted IMF (see e.g. Tables~1 and 2). As a result, for 
   each IMF choice the composite stellar population, determining the global 
   chemical properties of the Galaxy, is more or less enriched in stars 
   falling in a given mass range. Hence the model predictions on specific 
   quantities related to the given mass range vary when changing the IMF 
   prescriptions.

\begin{figure*}
   \centering
   \includegraphics[width=\textwidth]{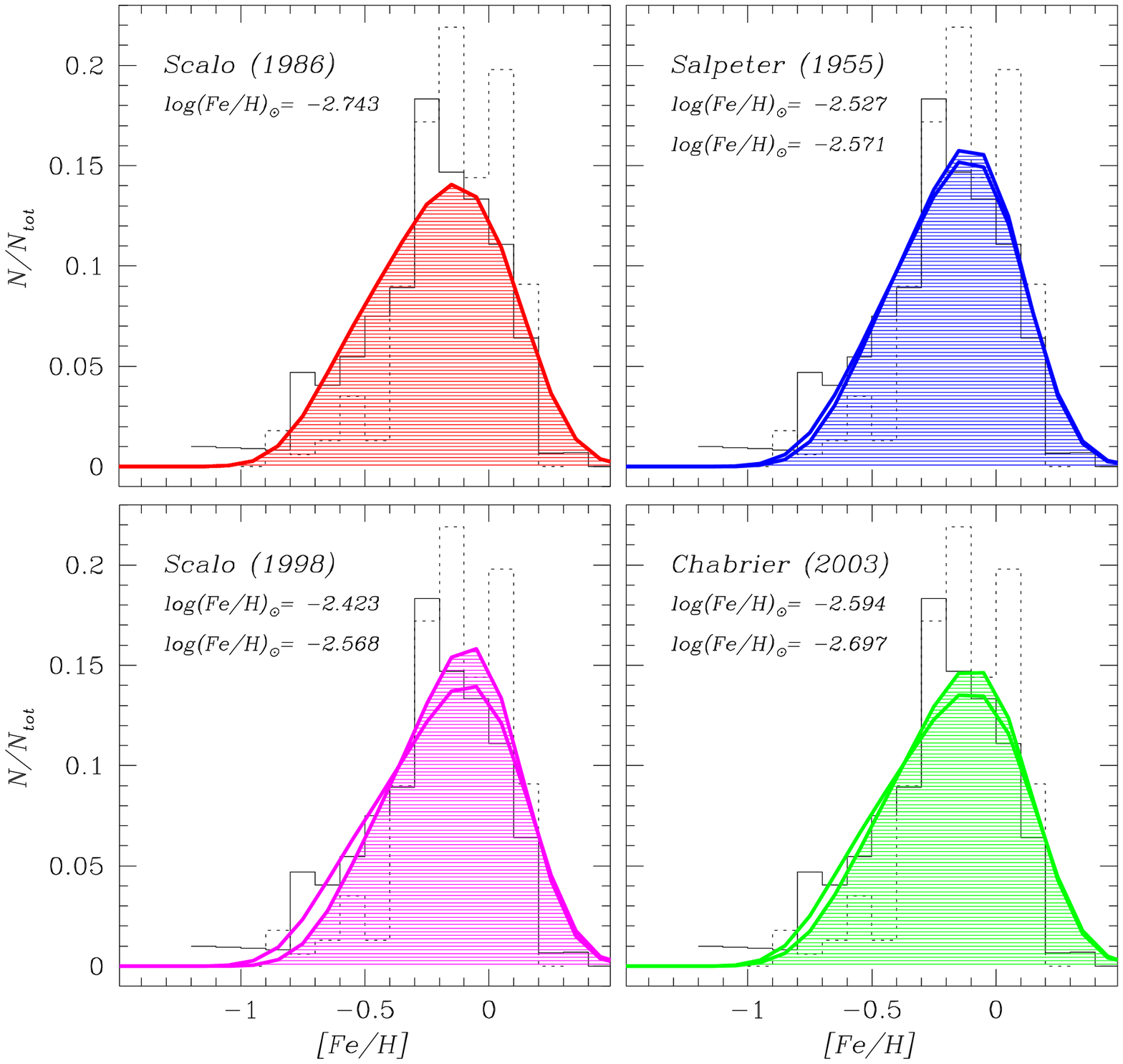}
      \caption{Theoretical G-dwarf metallicity distribution as predicted by 
               the model with Maeder \& Meynet (1989) stellar lifetimes and 
	       different assumptions on the IMF {\it (thick lines)} compared 
	       to the data ({\it histograms; solid:} Rocha-Pinto \& Maciel 
	       1996; {\it dotted:} J\o rgensen 2000). Theoretical [Fe/H] are 
	       normalized to the iron to hydrogen ratios predicted by the 
	       model at Sun's birth, which are displayed in the upper left 
	       corner of each panel. The first value refers to a common choice 
	       for the parameter regulating the fraction of stars giving rise 
	       to SNIa events, $A$ = 0.05; the second value refers to the case 
	       in which the $A$ parameter is adjusted so as to predict the 
	       same SNIa rate at the present time when changing the IMF (see 
	       text for details). The theoretical distributions corresponding 
	       to the latter case are shown as dashed areas. Remember that 
	       SNeIa are the major iron producers in the Milky Way. Notice 
	       that theoretical distributions are convolved with a Gaussian in 
	       order to account for both the observational and the intrinsic 
	       scatter.}
         \label{FigGDWARFS}
   \end{figure*}

   As a first example, Fig.~\ref{FigGDWARFS} displays the theoretical G-dwarf 
   metallicity distribution predicted by the model under different assumptions 
   on the IMF and with fixed stellar lifetimes (namely those by Maeder \& 
   Meynet 1989). The theoretical distributions are convolved with a Gaussian 
   to account for both the observational and the intrinsic scatter. The 
   adopted total dispersion is $\sigma$ = 0.15 dex in [Fe/H] (Arimoto, Yoshii 
   \& Takahara 1992; Kotoneva et al. 2002). The various panels show, clockwise 
   from top left, the theoretical distributions expected when assuming a Scalo 
   (1986), Salpeter (1955), Chabrier (2003) or Scalo (1998) IMF {\it (thick 
   lines)}. In particular, Chabrier's IMF has $x$ = 1.7 as the exponent of the 
   power-law for $m \ge$ 1 $M_\odot$. For all these IMFs, the normalization is 
   performed in the mass range 0.1--100 $M_\odot$. The distribution obtained 
   assuming Kroupa et al. IMF looks indistinguishable from that obtained 
   adopting Chabrier IMF. Similarly, the distribution obtained with Tinsley 
   IMF looks like that predicted with Scalo's (1998). Therefore, we do not 
   show them. The position of the peak and, in general, the shape of the 
   distribution are rather insensitive to the assumed IMF. They rather depend 
   on the adopted time scale of disk formation (e.g. Chiappini et al. 1997). 
   In particular, only a few stars are found at [Fe/H] $< -$0.7 dex, 
   independently of the choice of the IMF. Here we should point out that our 
   model predictions are for thin disk stars, whereas the observed metallicity 
   distribution cannot avoid some contribution from thick-disk stars, 
   especially at low metallicities.

%
%

   Also shown are data from Rocha-Pinto \& Maciel (1996; {\it thin solid 
   histograms}) and J\o rgensen (2000; {\it thin dotted histograms}). The two 
   distributions look quite different. In the case of J\o rgensen's, the peak 
   is shifted to higher metallicities and the low-metallicity tail is almost 
   absent, similarly to what found by Wyse \& Gilmore (1995). This is because 
   of the different metallicity calibrations and the different corrections to 
   the raw data applied by the authors. Indeed, in Rocha-Pinto \& Maciel 
   (1996) the metallicity scale is biased towards metal-poor stars (Martell \& 
   Laughlin 2002). This determines a shift of the peak of the distribution 
   towards lower metallicities. The actual peak position is likely to be more 
   near $-0.15$ dex than $-0.25$ dex (H. Rocha-Pinto, private communication), 
   thus bringing model predictions and observational distributions in better 
   agreement.

%
   \begin{table}
   \caption[]{Type Ia SN rates and Type Ia to Type II SN rate ratios at the 
              present time predicted by the model with different IMFs. All the 
	      IMFs are normalized to the mass range 0.1--100 $M_\odot$. The 
	      value of the parameter $A$, determining the stellar mass 
	      fraction belonging to binary systems with the right 
	      characteristics to give rise to SNIa explosions, is the same 
	      independently of the adopted IMF. Observed values are from van 
	      den Bergh \& Tammann (1991) and Cappellaro et al. (1997).}
   \begin{center}
   \begin{tabular}{c c c c}
   \hline \hline
   $\mathrm{IMF}$ & $A$ & $R_{\mathrm{Ia}} \; {\mathrm{(century^{-1})}}$ & 
   $R_{\mathrm{Ia}}/R_{\mathrm{II}}$ \\
   \hline
   ${\mathrm{S55}}$ & $0.05$ & $0.57$ & $0.34$ \\
   ${\mathrm{T80}}$ & $0.05$ & $1.16$ & $0.30$ \\
   ${\mathrm{S86}}$ & $0.05$ & $0.45$ & $0.46$ \\
   ${\mathrm{K93}}$ & $0.05$ & $0.67$ & $0.38$ \\
   ${\mathrm{S98}}$ & $0.05$ & $1.03$ & $0.26$ \\
   ${\mathrm{C03}}$ & $0.05$ & $0.69$ & $0.34$ \\
   \hline
   \multicolumn{2}{l}{${\mathrm{Observed}}$}
   & $0.6 \; h^2$ & $0.15$--$0.27$ \\
   \hline
   \end{tabular}
   \end{center}
   S55 -- Salpeter (1955); T80 -- Tinsley (1980); S86 -- Scalo (1986); K93 -- 
   Kroupa et al. (1993); S98 -- Scalo (1998); C03 -- Chabrier (2003) with $x$ 
   = 1.7.
   \end{table}
%

%
   \begin{table}
   \caption[]{Same as Table~3, but changing the value of $A$ to recover the 
              same SNIa rate at the present time with different IMFs.}
   \begin{center}
   \begin{tabular}{c c c c}
   \hline \hline
   $\mathrm{IMF}$ & $A$ & $R_{\mathrm{Ia}} \; {\mathrm{(century^{-1})}}$ & 
   $R_{\mathrm{Ia}}/R_{\mathrm{II}}$ \\
   \hline
   ${\mathrm{S55}}$ & $0.04$  & $0.47$ & $0.27$ \\
   ${\mathrm{T80}}$ & $0.02$  & $0.46$ & $0.17$ \\
   ${\mathrm{S86}}$ & $0.05$  & $0.45$ & $0.46$ \\
   ${\mathrm{K93}}$ & $0.035$ & $0.45$ & $0.25$ \\
   ${\mathrm{S98}}$ & $0.023$ & $0.47$ & $0.12$ \\
   ${\mathrm{C03}}$ & $0.032$ & $0.43$ & $0.23$ \\
   \hline
   \multicolumn{2}{l}{${\mathrm{Observed}}$}
   & $0.6 \; h^2$ & $0.15$--$0.27$ \\
   \hline
   \end{tabular}
   \end{center}
   S55 -- Salpeter (1955); T80 -- Tinsley (1980); S86 -- Scalo (1986); K93 -- 
   Kroupa et al. (1993); S98 -- Scalo (1998); C03 -- Chabrier (2003) with $x$ 
   = 1.7.
   \end{table}
%

   It is worth reminding that iron originates mostly from SNIa explosions 
   occurring in binary systems with an intermediate-mass primary. Only one 
   third of its solar abundance is related to SNII events, occurring on much 
   shorter time scales. Therefore, the stars responsible for the observed 
   behaviour of the iron abundance belong mostly to the 1.5--8 $M_\odot$ 
   stellar mass range, that is the one characterizing Type Ia SN progenitors 
   in our model.

   For each stellar generation, the mass fraction in form of binaries having 
   the right characteristics to end up as SNeIa is a free parameter of the 
   model, whose value is kept constant in time. Following Matteucci \& Greggio 
   (1986), the rate of Type Ia SNe is:
   \begin{displaymath}
     R_{\mathrm{Ia}}(t) = A 
     \int_{m_{B_{\mathrm{m}}}}^{m_{B_{\mathrm{M}}}} \varphi(m_B) \bigg[ 
     \int_{\mu_{\mathrm{m}}}^{0.5} f(\mu) \psi(t - \tau_{m_2})
     {\mathrm{d}}\mu \bigg] {\mathrm{d}}m_B,
   \end{displaymath}
   where $\varphi(m_B)$ is the IMF for the total mass of the binary system; 
   $f(\mu)$ is the distribution function for the mass fraction of the 
   secondary component ($\mu$ = $m_2/m_B$); $\tau_{m_2}$ is the lifetime of 
   the secondary; $m_{B_{\mathrm{m}}}$ = 3 $M_\odot$ and $m_{B_{\mathrm{M}}}$ 
   = 16 $M_\odot$ (see Matteucci \& Recchi 2001 for a review and alternative 
   formulations). The parameter $A$ is fixed by the request of reproducing the 
   SNIa rate currently observed in the disk. Moreover, the ratio 
   $R_{\mathrm{Ia}}/R_{\mathrm{II}}$ between the Type Ia and II SN rates in 
   the solar vicinity should be reproduced.

%
   \begin{table*}
   \caption[]{Abundances in the Protosolar Cloud as predicted by the model at 
              $t$ = 9.2 Gyr with different assumptions on the IMF. The 
              abundances are by number in $\log \varepsilon$(X), except for 
	      helium ($Y$) and global metallicity ($Z$), for which the 
	      abundances by mass are reported. Model predictions are compared 
              to observed photospheric solar abundances, unless otherwise 
	      stated.}
   \begin{center}
   \scriptsize{
   \begin{tabular}{@{} c c c c c c c c c c c @{}}
   \hline \hline
   $\mathrm{IMF}$ & $Y$ & ${\mathrm C}$ & ${\mathrm N}$ & ${\mathrm O}$ & 
   ${\mathrm{Ne}}$ & ${\mathrm{Mg}}$ & ${\mathrm{Si}}$ & ${\mathrm S}$ & 
   ${\mathrm{Fe}}$ & $Z$ \\
   \hline
   ${\mathrm{S55a}}$ & $0.264$ & $8.53$ & $8.33$ & $9.12$ & $8.27$ & $7.55$ & 
   $7.81$ & $7.39$ & $7.68$ & $0.028$ \\
   ${\mathrm{T80a}}$ & $0.273$ & $8.56$ & $8.46$ & $8.98$ & $8.07$ & $7.39$ & 
   $7.82$ & $7.40$ & $7.71$ & $0.024$ \\
   ${\mathrm{S86a}}$ & $0.260$ & $8.36$ & $8.04$ & $8.77$ & $7.84$ & $7.14$ & 
   $7.52$ & $7.11$ & $7.51$ & $0.014$ \\
   ${\mathrm{K93a}}$ & $0.264$ & $8.46$ & $8.18$ & $8.87$ & $7.97$ & $7.27$ & 
   $7.62$ & $7.21$ & $7.55$ & $0.017$ \\
   ${\mathrm{S98a}}$ & $0.271$ & $8.64$ & $8.49$ & $9.18$ & $8.36$ & $7.64$ & 
   $7.87$ & $7.44$ & $7.68$ & $0.033$ \\
   ${\mathrm{C03a}}$ & $0.265$ & $8.48$ & $8.21$ & $8.89$ & $8.00$ & $7.30$ & 
   $7.64$ & $7.23$ & $7.55$ & $0.018$ \\
   ${\mathrm{C03b}}$ & $0.263$ & $8.44$ & $8.15$ & $8.85$ & $7.95$ & $7.25$ & 
   $7.60$ & $7.19$ & $7.54$ & $0.017$ \\
   \cline{1-11}
   \multicolumn{11}{c}{${\mathrm{Observations}}$} \\
   \cline{1-11}
   ${\mathrm{GS98}}$ & $0.275 \pm 0.01$ & $8.52 \pm 0.06$ & $7.92 \pm 0.06$ & 
   $8.83 \pm 0.06$ & $8.08 \pm 0.06^{\mathrm{a}}$ & $7.58 \pm 0.05$ & $7.55 
   \pm 0.05$ & $7.33 \pm 0.11$ & $7.50 \pm 0.05$ & $0.017$ \\
   ${\mathrm{H01}}$ & $$ & $8.59 \pm 0.11$ & $7.93 \pm 0.11$ & $8.74 \pm 0.08$ 
   & $$ & $7.54 \pm 0.06$ & $7.54 \pm 0.05$ & $$ & $7.45 \pm 0.08$ & $$ \\
   ${\mathrm{AP01}}$ & $$ & $$ & $$ & $8.69 \pm 0.05$ & $$ & $$ & $$ & $$ & $$ 
   & $$ \\
   ${\mathrm{AP02}}$ & $$ & $8.39 \pm 0.04$ & $$ & $$ & $$ & $$ & $$ & $$ & $$ 
   & $$ \\
   ${\mathrm{A04}}$ & $$ & $$ & $$ & $8.66 \pm 0.05$ & $7.84 \pm 0.06$ & $$ & 
   $$ & $$ & $$ & $0.0126$ \\
   \hline
   \end{tabular}
   }
   \end{center}
   \scriptsize{
   \begin{list}{}{}
   \item[$^{\mathrm{a}}$] Coronal data.
   \end{list}
   S55 -- Salpeter (1955); T80 -- Tinsley (1980); S86 -- Scalo (1986); K93 -- 
   Kroupa et al. (1993); S98 -- Scalo (1988); C03 -- Chabrier (2003) with $x$ 
   = 1.7. The letter refers to the adopted normalization range: a stands for 
   0.1--100 $M_\odot$; b for 0.001--100 $M_\odot$. GS98 -- Grevesse \& Sauval 
   (1998); H01 -- Holweger (2001); AP01 -- Allende Prieto et al. (2001); AP02 
   -- Allende Prieto et al. (2002); A04 -- Asplund et al. (2004).
   }
   \end{table*}
%
  
   In Table~3 we list the present-day values of $R_{\mathrm{Ia}}$ and 
   $R_{\mathrm{Ia}}/R_{\mathrm{II}}$ which are predicted by assuming different 
   IMFs, while keeping the value of $A$ constant (the often quoted value of 
   0.05 for $A$ is adopted). If $h$ = 0.7 and a factor of two of uncertainty 
   in the observations is assumed, it turns out that both a Tinsley (1980) and 
   a Scalo (1998) IMF provide only a marginal agreement with the observed SNIa 
   rate. The reason for this is that, among all the studied IMFs, Tinsley's 
   (1980) and Scalo's (1998) are those allocating the highest mass fraction to 
   the stellar mass range 1.5--8 $M_\odot$, typical of SNIa progenitors.

   Table~4 illustrates what happens if the value of $A$ is adjusted so as to 
   bring the value of $R_{\mathrm{Ia}}$ in agreement with the observations 
   when changing the IMF. The corresponding G-dwarf metallicity distributions 
   are shown in Fig.~\ref{FigGDWARFS} {\it (dashed areas)}. Lowering $A$ from 
   0.05 to 0.04, 0.023 and 0.032, in case of Salpeter (1955), Scalo (1998) and 
   Chabrier (2003) IMF, respectively, results in narrower, more peaked 
   distributions. Nevertheless, after convolution with a Gaussian accounting 
   for the observational and intrinsic error, the differences almost 
   completely level off. The agreement with the observations is always 
   satisfactory. The Chabrier IMF also gives a good fit to the 
   $R_{\mathrm{Ia}}/R_{\mathrm{II}}$ ratio currently observed in the solar 
   vicinity. For the Scalo (1986) IMF, a better value for the 
   $R_{\mathrm{Ia}}/R_{\mathrm{II}}$ ratio can be obtained by further reducing 
   the $A$ value. The Kroupa et al. (1993) IMF is an equally valid choice. 
   However, no firm conclusions can be drawn on the grounds of the SN rates, 
   given the high uncertainty level still affecting the data. In what follows, 
   for each IMF we discuss the results obtained with the corresponding $A$ 
   value listed in Table~4, unless otherwise specified.

%
   \begin{figure*}
   \centering
   \includegraphics[width=17.2cm]{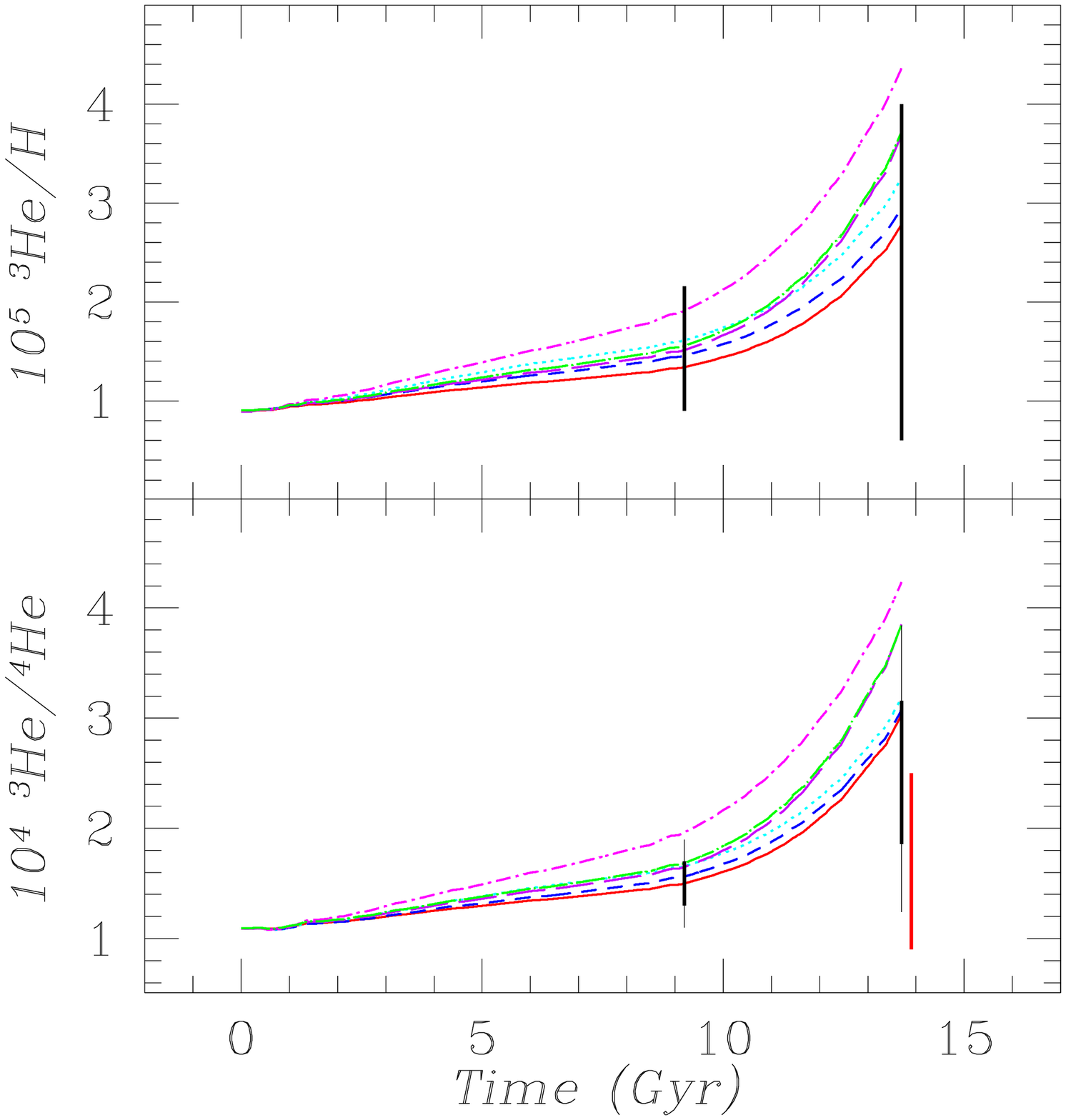}
      \caption{{\it Upper panel:} evolution of $^3$He/H in the solar 
               neighbourhood. Different lines refer to different IMF 
	       prescriptions: {\it short-dashed line:} Salpeter (1955); {\it 
               dotted line:} Tinsley (1980); {\it continuous line:} Scalo 
	       (1986); {\it long-dashed line:} Kroupa et al. (1993); {\it 
               dot-short-dashed line:} Scalo (1998); {\it dot-long-dashed 
	       line:} Chabrier (2003) with $x$ = 1.7 for $m >$ 1 $M_\odot$. 
               Data (vertical bars at $t$ = 9.2 and 13.7 Gyr) are from Geiss 
	       \& Gloeckler (1998). {\it Lower panel:} evolution of 
	       $^3$He/$^4$He in the solar neighbourhood. Models are labeled as 
	       in the upper panel. Data are from Geiss \& Gloeckler (1998; at 
	       1 and 2\,$\sigma$ -- thick and thin vertical bars, 
	       respectively, at $t$ = 9.2 and 13.7 Gyr) and Salerno et al. 
	       (2003; 1\,$\sigma$-bar on the right at $t$ = 13.7 Gyr).
	       }
         \label{FigHE}
   \end{figure*}
   \begin{figure*}
   \centering
   \includegraphics[width=17.2cm]{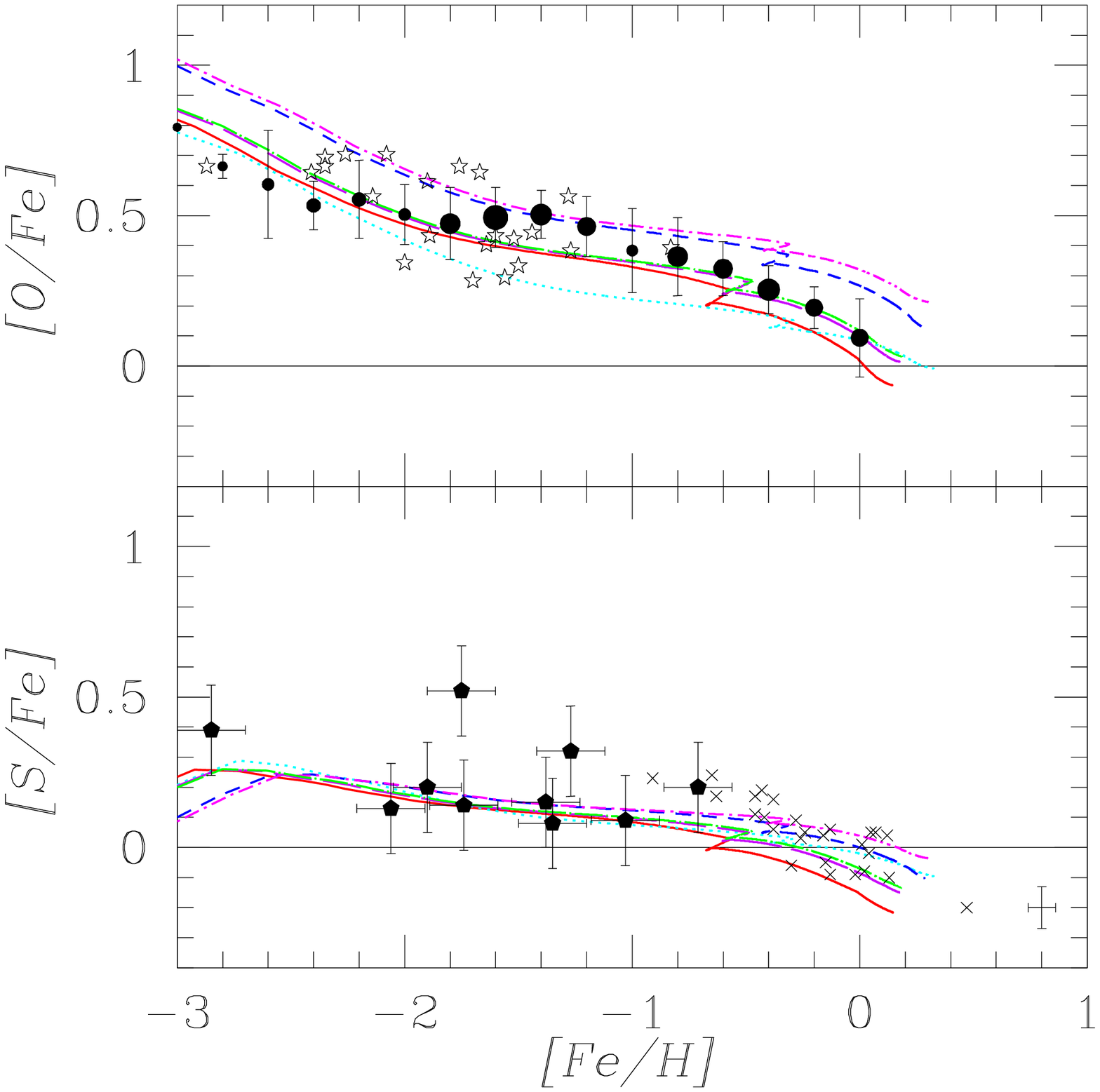}
      \caption{[O/Fe] {\it (upper panel)} and [S/Fe] {\it (lower panel)} 
	       ratios as functions of [Fe/H]. Models as in Fig.~\ref{FigHE}. 
	       All the ratios are normalized to solar elemental abundances by 
	       Grevesse \& Sauval (1998), except for oxygen, for which the 
	       solar value from Holweger (2001) is adopted. Data for oxygen 
	       are from Mel\'endez \& Barbuy (2002; {\it stars:} [O/Fe] values 
	       from infrared OH lines; {\it full circles:} mean [O/Fe] from 
	       [O\,{\small{I}}] line in bins of 0.2 dex in [Fe/H]; the size of 
	       the circles represents the number of stars in each metallicity 
	       bin). Data for sulphur are from Ryde \& Lambert (2004; {\it 
	       pentagons:} measurements from observations of S\,{\small{I}} 
	       lines at 9212.9, 9228.1 and 9237.5 \AA) and Chen et al. (2002; 
	       {\it crosses}). For these latter measurements, the typical 
	       error is also reported on the lower right corner of the panel.
               }
         \label{FigOS}
   \end{figure*}
%

   In Table~5 we display the abundances of $^4$He, C, N, O, Ne, Mg, Si, S, Fe 
   and the global metallicity, $Z$, predicted by the model at the time of the 
   Solar System formation 4.5 Gyr ago, under different prescriptions on the 
   IMF. Theoretical values are compared to observed ones from Grevesse \& 
   Sauval (1998), Holweger (2001), Allende Prieto, Lambert \& Asplund (2001, 
   2002) and Asplund et al. (2004). One may wonder whether the use of the Sun 
   as representative of the chemical composition of the ISM of the local disc 
   4.5 Gyr ago is suitable for comparison with GCE model predictions. Indeed, 
   the empirical age-metallicity relationship for solar neighbourhood stars 
   shows a large dispersion (Edvardsson et al. 1993), with known 
   planet-harbouring stars being systematically more metal rich than stars 
   without planets (Ibukiyama \& Arimoto 2002, their figure~21). On the other 
   hand, a recent reappraisal of the chemical composition of the Orion nebula 
   suggests heavy element abundances only slightly higher ($\sim$ 0.1 dex) 
   than the solar ones, in agreement with GCE model predictions, challenging 
   the view that the Sun has abnormally high metal abundances (Esteban et al. 
   2004).

   It is seen that a model adopting Tinsley (1980) or Scalo (1998) IMF is in 
   better agreement with the solar helium mass fraction (see also Romano et 
   al. 2003), but overestimates the overall metallicity of the gas and the 
   iron content. This is due to the high stellar mass fraction distributed 
   over the mass range 2--40 $M_\odot$ according to these IMFs (see Table~1), 
   coupled with the high helium and global metal yields predicted in this mass 
   range by stellar evolution theory (see figures 2.9, 2.10 of Matteucci 
   2001). Trying to reduce the predicted solar iron abundance by further 
   lowering the $A$ parameter is not a viable solution (although at first 
   glance it could appear as a very simple and promising one), since a too low 
   $R_{\mathrm{Ia}}/R_{\mathrm{II}}$ ratio would be obtained in this case, in 
   disagreement with the available data. Notice that a better agreement with 
   the $^4$He data can be achieved by model S86a if adopting the $^4$He yields 
   recently computed by Meynet \& Maeder (2002), which include mass loss and 
   rotation in self-consistent stellar evolutionary models. Indeed, a value of 
   0.272 for $Y$ at the time of Sun formation is predicted in this case, in 
   very good agreement with the observed solar value (Chiappini, Matteucci \& 
   Meynet 2003b).

   Salpeter's IMF also predicts far too high values for the solar abundances 
   of iron, oxygen and metals in general, because of its high percentage of 
   massive stars (see also Tosi 1988, 1996). In particular, with this IMF and 
   the star formation law adopted here, there happens to be no way to 
   reconcile the theoretical solar abundances with the observed ones, unless 
   one requires a SF process so unefficient that severe discrepancies with 
   other observational constraints do arise. In particular, by using $\nu 
   \sim$ 0.2 Gyr$^{-1}$ in the SFR expression reported in Section~4, the model 
   matches the actual metallicity of the Sun, but underestimates the 
   present-day stellar mass density in the solar vicinity. In fact, in this 
   case the expected stellar mass density at the present time turns out to be 
   $\Sigma_{\mathrm{stars}}(t_{\mathrm{Gal}}) \simeq$ 22 $M_\odot$ pc$^{-2}$, 
   to be compared with an observed value of 35 $\pm$ 5 $M_\odot$ pc$^{-2}$.

   The nitrogen solar abundance is overestimated by all the models, except by 
   the model assuming Scalo's (1986) IMF. This same model also predicts a 
   solar carbon abundance in good agreement with the observations, if one 
   believes that the solar photospheric abundance from Allende Prieto et al. 
   (2002) gives the best estimate of carbon in the Sun. A model assuming 
   Kroupa et al. or Chabrier's IMF also agrees with the CNO data, but only 
   at the 2-$\sigma$ level. However, it should be noticed that: (i) the 
   nitrogen and carbon yields from low- and intermediate-mass stars are very 
   uncertain and (ii) the carbon yields from stars with $m >$ 40 $M_\odot$ 
   have been multiplied by a factor of three in order to mimic results from 
   recent models taking mass loss and stellar rotation into account (see 
   Chiappini et al. 2003a, b and references therein) and both of these 
   mechanisms are still far from being fully understood. In conclusion, it is 
   likely that the uncertainties in the stellar nucleosynthesis are the most 
   important sources of errors as far as both carbon and nitrogen evolution 
   are concerned.

   The effect of extending the lognormal form of the IMF derived by Chabrier 
   (2003) for $m <$ 1 $M_\odot$ to the BD domain is also studied. BDs are 
   low-mass, long-living objects, that act simply as sinks of matter from the 
   point of view of galactic chemical evolution. They were born at every time 
   of Galaxy's evolution and never restored matter into the ISM. According to 
   what reported in Table~2, only a small fraction ($<$ 10\%) of the total 
   mass of a stellar generation should be locked up in BDs. Therefore, only 
   small effects are expected to appear when extending the IMF to the BD 
   domain. This is what is actually found (compare model C03a to model C03b 
   results in Table~5). Nevertheless, considering the existence of a 
   substellar mass regime reduces the amount of matter processed through 
   nuclear burning in stars during the whole Galactic lifetime. This results 
   in lower abundances in the gaseous matter at the time of Sun's formation, 
   thus obtaining a better agreement with the observations. The current 
   stellar density we obtain in the solar neighbourhood in the two cases is 
   fairly similar, being $\Sigma_{\mathrm{stars}}(t_{\mathrm{Gal}}) \simeq$ 34
   $M_\odot$ pc$^{-2}$ if BDs are not taken into account and 
   $\Sigma_{\mathrm{stars}}(t_{\mathrm{Gal}}) \simeq$ 29 $M_\odot$ pc$^{-2}$ 
   if they are. Both values are within the errors associated with the 
   observational estimate, though the higher value is more likely.

   Figs.~\ref{FigHE} and \ref{FigOS} show the errors associated with chemical 
   evolution model predictions for several quantities, obtained by changing 
   the prescriptions for the stellar IMF and keeping the same stellar 
   lifetimes (Maeder \& Meynet 1989). Model predictions on the evolution of 
   $^3$He/H and $^3$He/$^4$He as functions of time (Fig.~\ref{FigHE}, {\it 
   upper} and {\it lower panel,} respectively) and [O/Fe] and [S/Fe] as 
   functions of [Fe/H] (Fig.~\ref{FigOS}, {\it upper} and {\it lower panel,} 
   respectively) in the solar neighbourhood are shown, as well as the 
   corresponding data. Model predictions are not normalized to the predicted 
   solar values, as instead is often done in chemical evolution studies. This 
   is to better appreciate the differences obtained with the various IMFs in 
   the last 4.5 Gyr of evolution, an effect which is not evident when the 
   solar normalization is applied. The elements displayed in Figs.~\ref{FigHE} 
   and \ref{FigOS} are not chosen randomly. They are representative of stellar 
   progenitors belonging to different initial mass ranges: (i) $^3$He is 
   produced by stars belonging to the low-mass range, 1--2 $M_\odot$; (ii) 
   $^4$He comes from the whole stellar mass range; (iii) oxygen is almost 
   entirely produced on short time scales by stars with $m >$ 10 $M_\odot$; 
   (iv) sulphur is representative of elements produced partly by SNeII and 
   partly by SNeIa; finally, (v) Fe is mostly produced by Type Ia SNe on long 
   timescales.

%
   \begin{figure}
   \centering
   \includegraphics[width=\columnwidth]{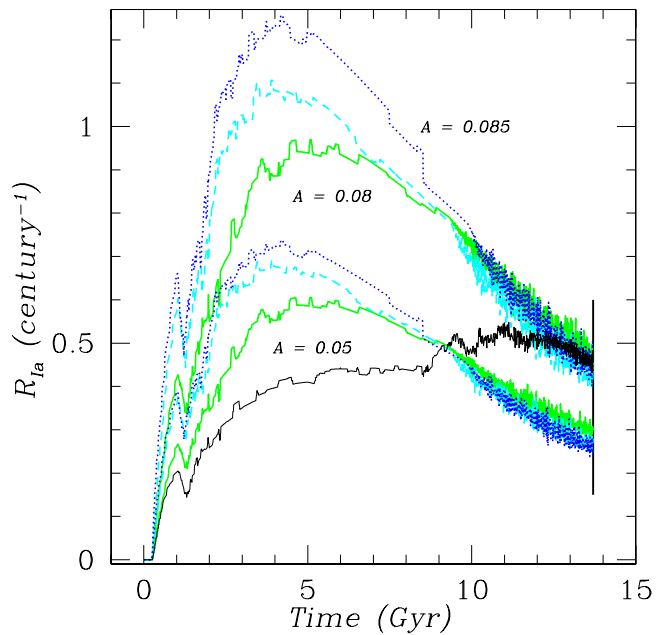}
      \caption{Type Ia SN rates obtained with different assumptions on the 
	       stellar lifetimes (Maeder \& Meynet 1989 -- {\it thin solid 
	       line}; Tinsley 1980 -- {\it thick dotted line}; Schaller et al. 
	       1992 -- {\it thick solid line}; Kodama 1997 -- {\it thick 
	       dashed line}). We also show the effect of changing the fraction 
	       of mass entering the formation of Type Ia SN progenitors. The 
	       four curves lying in the lower part of the diagram have all 
	       been computed with $A$ = 0.05, while the value of $A$ for the 
	       curves lying in the upper part of the diagram has been changed 
	       so as to produce the same present-day $R_{\mathrm{Ia}}$ in the 
	       disk, independently of the adopted stellar lifetime 
	       prescriptions. The Type Ia SN rate observed in the Galaxy at 
	       the present time is also shown ({\it vertical bar,} for $h$ = 
	       0.7; Cappellaro et al. 1997).
	      }
         \label{FigIaRATE}
   \end{figure}
%

%
   \begin{table*}
   \caption[]{Abundances in the PSC as predicted by the model at $t$ = 9.2 Gyr 
              for different assumptions on the stellar lifetimes. The 
              abundances are by number in $\log \varepsilon$(X), except for 
	      helium ($Y$) and global metallicity ($Z$), for which the 
	      abundances by mass are reported. Model predictions are compared 
              to observed photospheric solar abundances, unless otherwise 
	      stated. A value of $A$ = 0.05 is assumed for all models (see 
	      text).}
   \begin{center}
   \scriptsize{
   \begin{tabular}{@{} c c c c c c c c c c c @{}}
   \hline \hline
   $\tau_m$ & $Y$ & ${\mathrm C}$ & ${\mathrm N}$ & ${\mathrm O}$ & 
   ${\mathrm{Ne}}$ & ${\mathrm{Mg}}$ & ${\mathrm{Si}}$ & ${\mathrm S}$ & 
   ${\mathrm{Fe}}$ & $Z$ \\
   \hline
   ${\mathrm{T80}}$ & $0.261$ & $8.39$ & $8.13$ & $8.90$ & $7.97$ & $7.23$ & 
   $7.57$ & $7.18$ & $7.57$ & $0.017$ \\
   ${\mathrm{M89}}$ & $0.260$ & $8.36$ & $8.04$ & $8.77$ & $7.84$ & $7.14$ & 
   $7.52$ & $7.11$ & $7.51$ & $0.014$ \\
   ${\mathrm{S92}}$ & $0.260$ & $8.36$ & $8.06$ & $8.80$ & $7.89$ & $7.17$ & 
   $7.55$ & $7.10$ & $7.54$ & $0.015$ \\
   ${\mathrm{K97}}$ & $0.260$ & $8.35$ & $8.08$ & $8.81$ & $7.89$ & $7.17$ & 
   $7.51$ & $7.11$ & $7.56$ & $0.015$ \\
   \cline{1-11}
   \multicolumn{11}{c}{${\mathrm{Observations}}$} \\
   \cline{1-11}
   ${\mathrm{GS98}}$ & $0.275 \pm 0.01$ & $8.52 \pm 0.06$ & $7.92 \pm 0.06$ & 
   $8.83 \pm 0.06$ & $8.08 \pm 0.06^{\mathrm{a}}$ & $7.58 \pm 0.05$ & $7.55 
   \pm 0.05$ & $7.33 \pm 0.11$ & $7.50 \pm 0.05$ & $0.017$ \\
   ${\mathrm{H01}}$ & $$ & $8.59 \pm 0.11$ & $7.93 \pm 0.11$ & $8.74 \pm 0.08$ 
   & $$ & $7.54 \pm 0.06$ & $7.54 \pm 0.05$ & $$ & $7.45 \pm 0.08$ & $$ \\
   ${\mathrm{AP01}}$ & $$ & $$ & $$ & $8.69 \pm 0.05$ & $$ & $$ & $$ & $$ & $$ 
   & $$ \\
   ${\mathrm{AP02}}$ & $$ & $8.39 \pm 0.04$ & $$ & $$ & $$ & $$ & $$ & $$ & $$ 
   & $$ \\
   ${\mathrm{A04}}$ & $$ & $$ & $$ & $8.66 \pm 0.05$ & $7.84 \pm 0.06$ & $$ & 
   $$ & $$ & $$ & $0.0126$ \\
   \hline
   \end{tabular}
   }
   \end{center}
   \scriptsize{
   \begin{list}{}{}
   \item[$^{\mathrm{a}}$] Coronal data.
   \end{list}
   T80 -- Tinsley (1980); M89 -- Maeder \& Meynet (1989); S92 -- Schaller et 
   al. (1992); K97 -- Kodama (1997); GS98 -- Grevesse \& Sauval (1998); H01 -- 
   Holweger (2001); AP01 -- Allende Prieto et al. (2001); AP02 -- Allende 
   Prieto et al. (2002); A04 -- Asplund et al. (2004).
   }
   \end{table*}
%

   Scalo's (1998) IMF, having the highest fractions of stars with $m >$ 10 
   $M_\odot$ and 1 $\le m/M_\odot \le$ 2 (Fig.~\ref{FigIMF1}), predicts a too 
   large O/Fe ratio during the whole Galactic lifetime and an increase of 
   $^3$He/H and $^3$He/$^4$He from the time of Sun's formation up to now that 
   hardly fits the data. This is because the stars in the range 1--2 $M_\odot$ 
   are net $^3$He producers, even when $\sim$90\% of them are assumed to 
   experience the cool bottom processes which strongly deplete their $^3$He 
   yields (see next section for more details and references), whereas 
   high-mass stars produce almost all the oxygen. On the contrary, Salpeter, 
   Tinsley and Scalo (1986) IMFs predict enrichment histories for $^3$He/H and 
   $^3$He/$^4$He which better agree with the data, because of the lower mass 
   fraction falling in the 1--2 $M_\odot$ stellar mass range. The Kroupa et 
   al. and Chabrier IMFs, with $x$ = 1.7 for $m >$ 1 $M_\odot$, produce an 
   intermediate behaviour. They also guarantee the best fit to the [O/Fe] vs 
   [Fe/H] data. The (small) differences in the S/Fe ratio in the disc are 
   mostly related to the (slightly) different behaviour of the SNIa rate in 
   the past predicted when adopting different IMFs.

   \subsection{Changing the stellar lifetimes}

   In this section we keep fixed the prescriptions on the stellar IMF, while 
   varying those on the stellar lifetimes, according to what has been reviewed 
   in Section~3. We choose to adopt Scalo (1986) IMF for the sake of 
   comparison with previous results published in a series of papers dealing 
   with different aspects of the Milky Way evolution (e.g. Chiappini et al. 
   1997; Romano et al. 2003, and references therein).

   In the following, we compare results of chemical evolution models for the 
   solar vicinity adopting stellar lifetimes from: (i) Tinsley (1980); (ii) 
   Maeder \& Meynet (1989)\footnote{In the $m \le$ 1.3 $M_\odot$, $m >$ 60 
   $M_\odot$ mass ranges we use the extrapolations adopted by Matteucci and 
   coworkers (Matteucci \& Fran\c cois 1989; Chiappini et al. 1997).}; (iii) 
   Schaller et al. (1992)\footnote{We use the polynomial fits of Gibson 
   (1997).}; (iv) Kodama (1997)\footnote{His prescriptions are equivalent to 
   those given in Padovani \& Matteucci (1993).}.

%
   \begin{figure*}
   \centering
   \includegraphics[width=\textwidth]{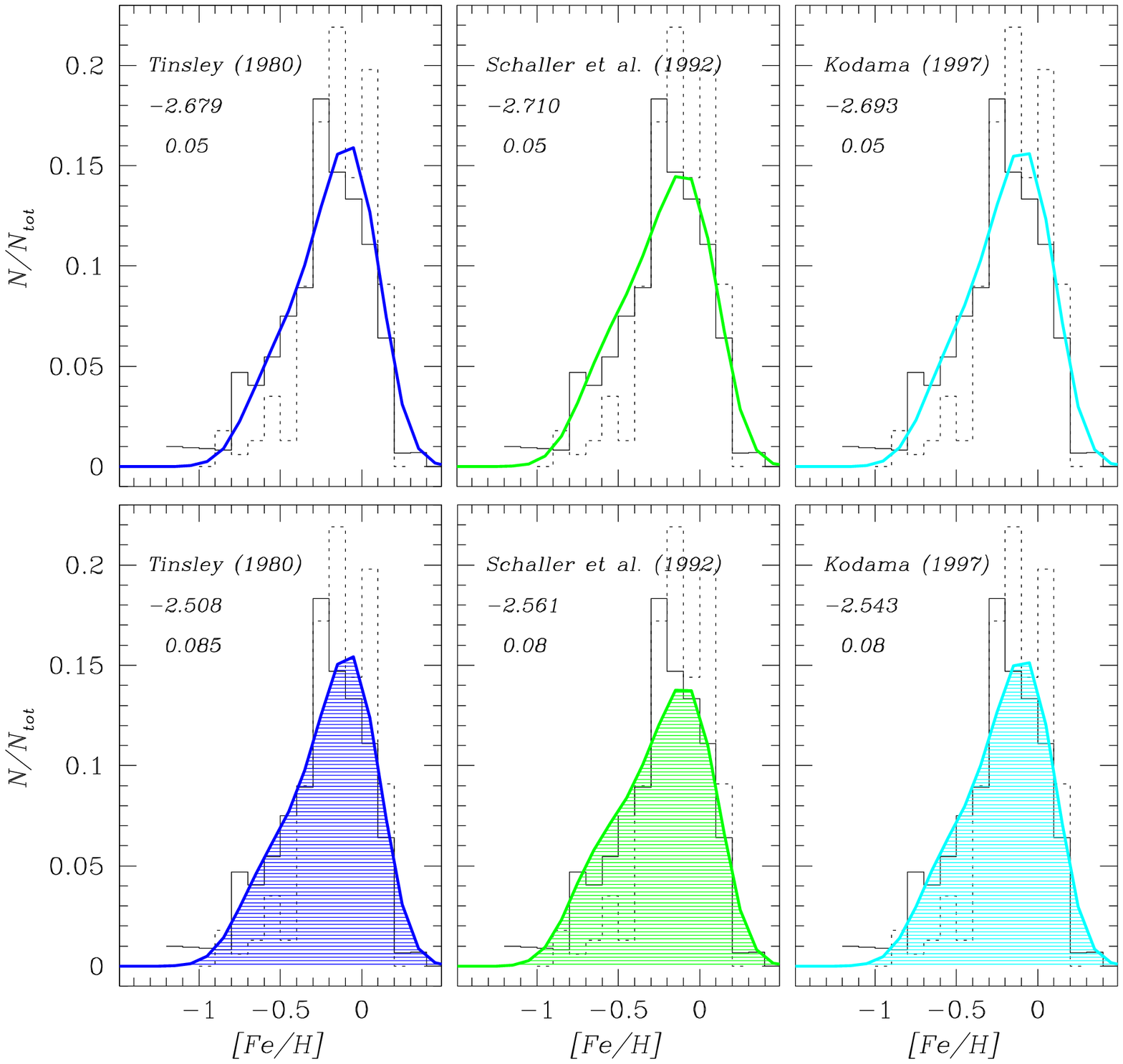}
      \caption{Theoretical G-dwarf metallicity distributions {\it (thick 
	       lines)} predicted by the model with different assumptions on 
	       the stellar lifetimes ({\it left panels:} Tinsley 1980; 
	       {\it middle panels:} Shaller et al. 1992; {\it right panels:} 
	       Kodama 1997) and on the fraction of mass entering the formation 
	       of Type Ia SN progenitors ({\it upper panels:} $A$ = 0.05; {\it 
	       lower panels:} $A$ = 0.08 in case of Schaller et al.'s and 
	       Kodama's stellar lifetimes; $A$ = 0.085 in case of Tinsley's 
	       stellar lifetimes). Scalo's (1986) IMF is assumed in all cases. 
	       Theoretical distributions are compared to observational 
	       ones from Rocha-Pinto \& Maciel (1996; {\it solid histograms}) 
	       and J\o rgensen (2000; {\it dotted histograms}). For each 
	       model, the theoretical [Fe/H] ratios are normalized to the Fe/H 
	       ratio at Sun's birth predicted by the model itself, given in 
	       the upper left corner of each panel (middle row). This produces 
	       a shift of the distribution along the x axis, thus allowing for 
	       a better comparison of the theoretical shape with the observed 
	       one even if the model does not recover the actual iron 
	       abundance of the Sun.
	      }
         \label{FigGD}
   \end{figure*}
%

   In Chiappini et al. (1997) and subsequent work by that group, the 
   prescriptions of Maeder \& Meynet (1989) were adopted. According to those 
   authors, stars with 1.3 $\le m/M_\odot \le$ 3 are characterized by 
   lifetimes longer than in any other study (see Fig.~\ref{FigTAU}). This 
   causes the Type Ia SN rate to increase smoothly during almost the whole 
   disk evolution, with a gentle decline starting only a couple of Gyrs ago 
   (thin solid line in Fig.~\ref{FigIaRATE}). On the contrary, models with 
   different prescriptions on the stellar lifetimes all produce a well-defined 
   peak in the rate, followed by a steep decline afterward (thick lines in 
   Fig.~\ref{FigIaRATE}). If one adopts Tinsley's (1980), Schaller et al.'s 
   (1992) or Kodama's (1997) stellar lifetimes, keeping $A$ -- i.e. the 
   fraction of mass belonging to binary systems which will give rise to Type 
   Ia SN explosions -- constant and equal to 0.05 results in a little bit more 
   iron produced until the time of Sun's formation (see also Table~6, column 
   10), while SNeIa are supposed to be less numerous in the disk at the 
   present time. However, their number still agrees with that inferred from 
   the observations (vertical bar in Fig.~\ref{FigIaRATE}). To rise the 
   present-day SNIa rate to the value expected when adopting Maeder \& Meynet 
   (1989) stellar lifetimes, it is necessary to increase the fraction of mass 
   belonging to SNIa progenitors. In particular, when adopting Schaller et al. 
   (1992) or Kodama (1997) $\tau_m$s, $A = 0.08$ is required, while a slightly 
   higher value, $A = 0.085$, must be assumed if Tinsley's (1980) stellar 
   lifetimes are adopted ({\it upper curves} in Fig.~\ref{FigIaRATE}). 
   Increasing the efficiency of star formation while leaving the $A$ parameter 
   unchanged would also lead to an increase of the SNIa rate, but it would 
   also cause major problems in reproducing other important constraints 
   available for the immediate solar vicinity. From Fig.~\ref{FigIaRATE} and 
   Table~6, column 10, we conclude that, for a Scalo (1986) IMF, a small $A$ 
   value of the order of 0.05 or even less should be preferred, because it 
   produces theoretical predictions matching remarkably well the solar iron 
   abundance and the present-day SNIa rate data at the same time.

   The G-dwarf metallicity distributions obtained with all the above-mentioned 
   prescriptions for $\tau_m$ are displayed in Figs.~\ref{FigGDWARFS} {\it 
   (upper left panel)} and \ref{FigGD}. Again, the theoretical distributions 
   are convolved with a Gaussian to account for both the observational and the 
   intrinsic scatter. The adopted total dispersion in [Fe/H] is $\sigma$ = 
   0.15 dex (Arimoto, Yoshii \& Takahara 1992; Kotoneva et al. 2002). Since 
   most of the iron in the solar neighbourhood comes from Type Ia SNe, the 
   shape of the distribution is expected to change according to changes 
   affecting Type Ia SN progenitors. Indeed, adopting Tinsley (1980) or Kodama 
   (1997) stellar lifetimes (Fig.~\ref{FigGD}, {\it left} and {\it right 
   panels,} respectively) results in a more pronounced peak, independently of 
   the value of $A$ (which is given in the upper left corner of each panel, 
   third row). 

%
   \begin{figure*}
   \centering
   \includegraphics[width=17.2cm]{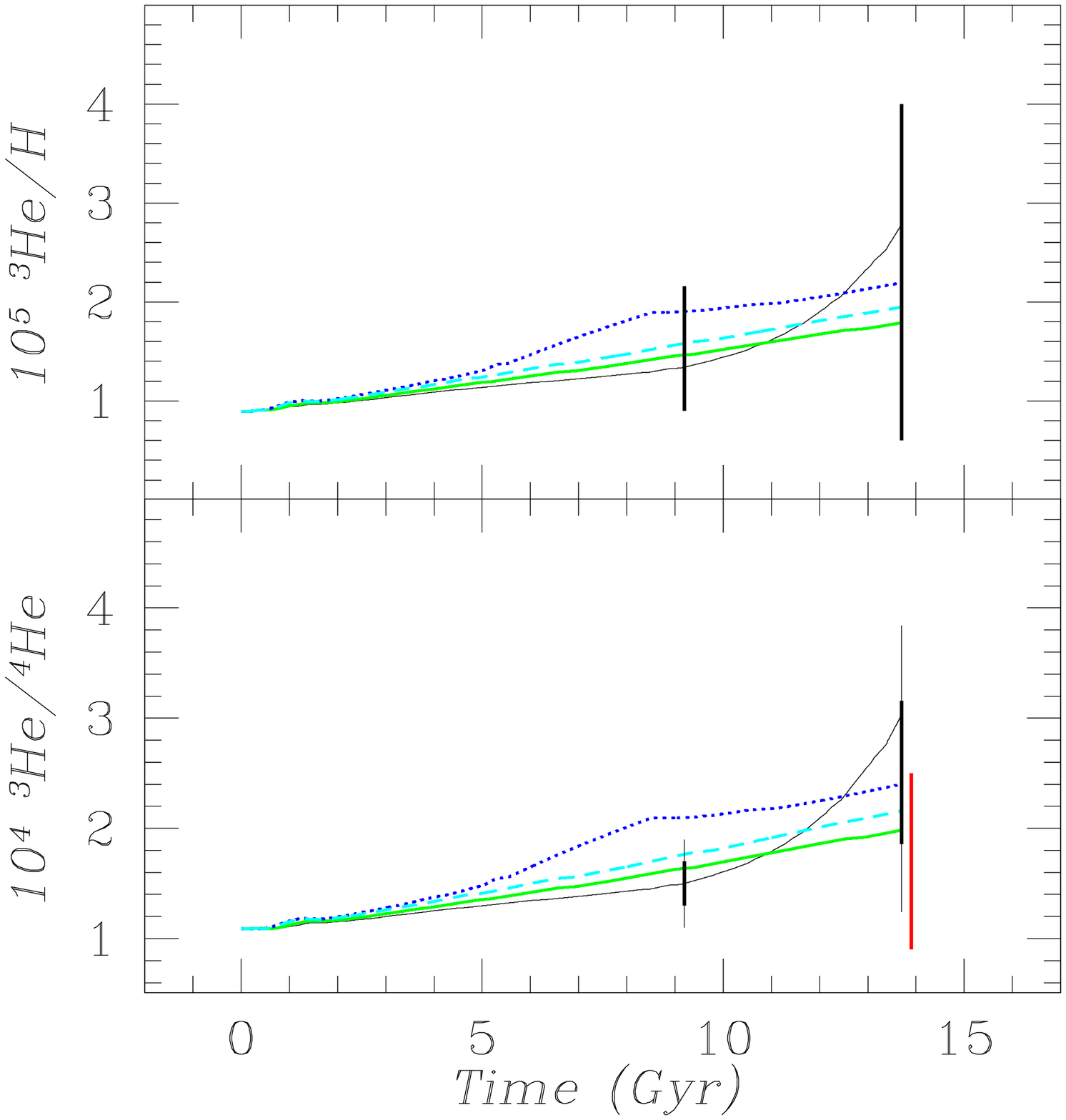}
      \caption{Same as Fig.~\ref{FigHE}, but here different line types refer 
	       to different prescriptions on the stellar lifetimes. {\it Thin 
	       solid lines:} Maeder \& Meynet (1989); {\it thick solid lines:} 
	       Schaller et al. (1992); {\it thick dotted lines:} Tinsley 
	       (1980); {\it thick dashed lines:} Kodama (1997). The Scalo 
	       (1986) IMF is assumed for all the models.
	      }
         \label{FigHETAU}
   \end{figure*}
%

   It is worth noticing that, independently of the choice of the stellar 
   lifetimes, we always get a strikingly good fit to the position of the peak 
   and to the high-metallicity tail of the distribution. We conclude that: (i) 
   the main parameter driving the location of the peak and the shape of the 
   distribution is the adopted time scale for thin-disk formation, which was 
   already well known (Chiappini et al. 1997); (ii) the adopted stellar 
   lifetimes affect the theoretical G-dwarf distribution as well, though 
   through a second order effect: they mostly act on the width and height of 
   the distribution. However, convolution with a Gaussian which accounts for 
   errors makes different distributions, obtained with different prescriptions 
   on the stellar lifetimes, look pretty much the same.

   In Table~6 we report the abundances in the Protosolar Cloud (PSC) predicted 
   under different assumptions on the stellar lifetimes. Notice that for all 
   the models in Table~6 a value of 0.05 for $A$ is adopted. A larger $A$ 
   value would overestimate the iron content of the Sun. It can be seen that 
   the global metallicity, $Z$, is well reproduced by all the models and so 
   are the CNO abundances, with a possible exception for oxygen, whose 
   predicted solar abundance is a bit higher than observed. On the other hand, 
   the helium abundance turns out to be too low. Increasing the star formation 
   efficiency would provide a higher $Y$ abundance, but the oxygen and global 
   metallicity would also increase, breaking the agreement with the 
   observations. This clearly shows that some revision on the helium stellar 
   yields is necessary (see Meynet \& Maeder 2002; Chiappini et al. 2003b).

   Let us now comment on specific trends predicted for a handful of important 
   species: (i) $^3$He and (ii) $^7$Li, coming mostly from low-mass stellar 
   progenitors; (iii) oxygen, a typical massive star product; and (iv) 
   sulphur, with both a SNII and a SNIa component. Major differences are 
   expected in the model predictions for $^3$He and $^7$Li, because of the 
   large differences characterizing different $\tau_m$ prescriptions for their 
   stellar progenitors.

   To deal with $^3$He, one must keep in mind that it is mostly produced by 
   low-mass -- hence long-living -- stars. Standard stellar evolutionary 
   theory predicts a large production of $^3$He from low-mass stars leading to 
   severe inconsistency between the observed $^3$He abundances and those 
   predicted by chemical evolution models assuming these standard yields 
   (Rood, Steigman \& Tinsley 1976; Galli et al. 1995; Olive et al. 1995; 
   Dearborn et al. 1996; Prantzos 1996). A problem that has been now 
   superseeded by the inclusion of rotational mixing in stellar models (e.g. 
   Charbonnel 1995; Sackmann \& Boothroyd 1999). Indeed, extra mixing makes it 
   possible to reconcile predictions from Galactic chemical evolution models 
   with observations, as long as $^3$He is destroyed in a large enough 
   fraction ($\sim$90\%) of low-mass stars (e.g. Galli et al. 1997; Chiappini 
   et al. 2002, and references therein). Fig.~\ref{FigHETAU} shows the $^3$He 
   evolution in the solar neighbourhood as predicted under different 
   assumptions on the stellar lifetimes: the {\it thin solid lines} are for 
   Maeder \& Meynet (1989); the {\it thick solid lines} are for Schaller et 
   al. (1992); the {\it thick dotted lines} for Tinsley (1980) and the {\it 
   thick dashed} ones for Kodama (1997). The prescriptions on $^3$He synthesis 
   are those from Dearborn et al. (1996) and Sackmann \& Boothroyd (1999) for 
   stars without and with extra-mixing. These $^3$He yields were recently 
   adopted also by Chiappini et al. (2002) and Romano et al. (2003) and take 
   extra-mixing in 93\% of low-mass stars into account. The lower $^3$He 
   production for $t <$ 11--12 Gyr predicted assuming Maeder \& Meynet (1989) 
   stellar lifetimes is due to the longer lifetimes in the mass range 1--2 
   $M_\odot$. The subsequent steep rise is mostly due to the late contribution 
   from stars in the 0.6--0.9 $M_\odot$ mass range, which die if adopting the 
   Matteucci \& Fran\c cois (1989) extrapolation of Maeder \& Meynet's stellar 
   lifetimes in the very low stellar mass range. Conversely, these stars never 
   die if the Schaller et al., Tinsley or Kodama stellar lifetimes are adopted 
   instead (Fig.~\ref{FigTAU}; see also the discussion in Romano et al. 2003). 
   If one trusts the local value of the helium isotopic ratio as measured with 
   the COLLISA experiment on board the space station {\it MIR} (Salerno et al. 
   2003 --  Fig.~\ref{FigHETAU}, {\it lower panel,} vertical bar on the right 
   at $t$ = 13.7 Gyr), longer lifetimes for stars with $m <$ 1 $M_\odot$ 
   should be preferred. In fact, this measurement supports the hypothesis that 
   negligible changes of the abundance of $^3$He occurred in the Galaxy during 
   the last 4.5 Gyr. 

%
   \begin{figure}
   \centering
   \includegraphics[width=\columnwidth]{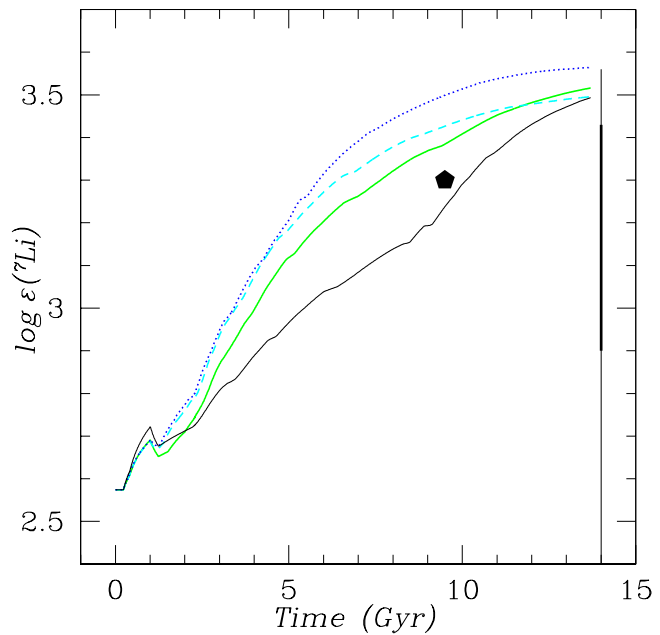}
      \caption{Temporal evolution of lithium in the solar neighbourhood. 
               Different curves refer to model predictions obtained with 
	       different recipes for the stellar lifetimes. {\it Thin solid 
	       line:} Maeder \& Meynet (1989); {\it thick solid line:} 
	       Schaller et al. (1992); {\it thick dotted line:} Tinsley (1980);
	       {\it thick dashed line:} Kodama (1997). The meteoritic 
	       (Nichiporuk \& Moore 1974; {\it pentagon}) and the depletion 
	       corrected value for the local ISM (Knauth et al. 2003; at 1- 
	       and 2-$\sigma$ -- {\it thick} and {\it thin vertical bars}, 
	       respectively) are shown as well. Notice that the $\alpha$ value 
	       (where $\alpha$ is the fraction of white dwarfs entering the 
	       formation of nova systems) is adjusted so as to reproduce the 
	       nova outburst rate currently observed in the Galaxy. In 
	       particular, in the case of Tinsley, Schaller et al. and Kodama 
	       stellar lifetimes $\alpha \sim$ 0.02, while $\alpha \sim$ 0.01 
	       if Maeder \& Meynet stellar lifetimes are adopted instead (see 
	       text for details). 
               }
         \label{FigLI}
   \end{figure}
%

   Lithium is another element originating mostly from low-mass stars (Romano 
   et al. 2001). According to our models, its meteoritic abundance is due for 
   $\sim$40\% to low-mass ($m$ = 1--2 $M_\odot$) stars on the red giant branch 
   (RGB) and for $\sim$10\% to novae, binary systems consisting of a white 
   dwarf plus a main sequence companion (Romano et al. 2001, 2003; but see 
   also Casuso \& Beckman 2000 and Travaglio et al. 2001 for different views). 
   The predicted steep rise of the lithium abundance in the ISM at late times 
   is dictated by the long time scales for lithium production from RGB stars 
   and novae when adopting Maeder \& Meynet stellar lifetimes. Here we analyse 
   what changes are introduced by adopting different stellar lifetimes. In 
   Fig.~\ref{FigLI}, we compare model predictions for the temporal evolution 
   of $^7$Li in the solar vicinity obtained by adopting Maeder \& Meynet 
   (1989) stellar lifetimes {\it (thin solid line)} to what we get if adopting 
   Schaller et al. (1992 -- {\it thick solid line}), Tinsley (1980 -- {\it 
   thick dotted line}) or Kodama (1997-- {\it thick dashed line}) 
   prescriptions. The Schaller et al. stellar lifetimes result in a more 
   gentle rise during the last $\sim$ 3 Gyrs, while almost no evolution is 
   expected during the same time interval with Tinsley (1980) or Kodama (1997) 
   prescriptions. These considerations read interesting in the light of recent 
   claims of a null $^7$Li evolution in the ISM during the last several Gyrs 
   inferred from cluster and field star lithium data (Lambert \& Reddy 2004). 
   However, in our opinion these data do not rule out a scenario of late 
   lithium pollution from low-mass stars, if the uncertainties in both the 
   models and the data are properly taken into account.

   In our model, nearly 10\% of lithium in meteorites is produced during nova 
   outbursts. These are periodical explosions, that do not destroy the parent 
   system. When dealing with nova systems, one must introduce in the model a 
   free parameter, $\alpha$, describing the fraction of white dwarfs which 
   enters the formation of nova systems (similarly to what is done for SNeIa; 
   see D'Antona \& Matteucci 1991; Romano et al. 1999; Matteucci et al. 2003). 
   Fig.~\ref{FigLI} shows model results obtained by changing the $\alpha$ 
   value so as to obtain the same theoretical nova outburst rate in the Galaxy 
   at the present time, whatever the $\tau_m$ choice. A value of 
   $R_{\mathrm{nova}}(t_{\mathrm{Gal}}) \simeq$ 20 yr$^{-1}$ is found with 
   $\alpha \sim 0.01$ in the case of Maeder \& Meynet stellar lifetimes and 
   $\alpha \sim 0.02$ in the remaining cases, to be compared with 
   $R_{\mathrm{nova}}^{\mathrm{obs}} =$ 20--30 yr$^{-1}$ (Shafter 1997). When 
   using Tinsley's, Schaller et al.'s or Kodama's stellar lifetimes, a $^7$Li 
   production from novae higher than expected when using Maeder \& Meynet's 
   stellar lifetimes is obtained during the whole Galactic lifetime, owing to 
   the higher value assumed for $\alpha$. Lowering the $^7$Li yields from red 
   giants and/or novae can make the theoretical predictions in better 
   agreement with the observations. This goes in the right direction. In fact, 
   in order to reproduce the observed meteoritic lithium data, in previous 
   work we had to require that all low-mass red giants are displaying 
   \emph{the largest observed atmospheric lithium enrichment} near the tip of 
   the RGB and couple this with \emph{the most efficient mass loss} still 
   compatible with observations (Romano et al. 2001). Adopting stellar 
   lifetimes prescritions different from Maeder \& Meynet (1989) helps us to 
   alleviate such an extreme, \emph{ad hoc} scenario.

   Finally, as far as the [O/Fe] and [S/Fe] ratios are concerned, it is worth 
   emphasizing that only small differences are found among models adopting 
   different stellar lifetimes. In the case of Tinsley's (1980) prescriptions 
   a flatter behaviour is predicted for [O/Fe] vs [Fe/H] at high 
   metallicities, at variance with observations (Bensby, Feltzing, \& 
   Lundstr\"om 2004). However, no firm conclusions can be drawn on this point 
   alone.

   \section{Final remarks and conclusions}

   Together with the time modulation of the SFR, the IMF dictates the 
   evolution and fate of galaxies. Nonluminous BDs and the lowest-mass stars 
   lock up an increasing fraction of the baryonic mass of galaxies over the 
   cosmological time. Short-lived massive stars and intermediate-mass stars 
   belonging to binary systems ending up as Type Ia SNe heat the ISM, 
   eventually determining the occurrence of galactic outflows. It is therefore 
   of much importance to quantify the effect of changing the relative numbers 
   of stars in different mass ranges at different times on galactic chemical 
   evolution model predictions.

   In this work we deal with our own Galaxy. We ascertain the range of 
   variations that affect several model predictions when accounting for 
   uncertainties on both the stellar IMF and the stellar lifetimes. First, we 
   show results of chemical evolution models for the Milky Way computed with 
   different assumptions on the stellar IMF. Then, we investigate the effects 
   of changing the prescriptions on the stellar lifetimes. `Theoretical 
   errors' are associated to model predictions due to uncertainties in the IMF 
   and/or stellar lifetimes. We summarize our main conclusions as follows:

   Among all the studied IMFs, the Salpeter (1955) and Scalo (1998) ones are 
   those in less agreement with the data. The Scalo (1986), Kroupa et al. 
   (1993) and Chabrier (2003) IMFs all guarantee the best fits to several 
   important observed properties of the solar vicinity. 

   Different stellar lifetime prescriptions differ mostly in the low and very 
   low stellar mass domain. Therefore, it is not surprising that models 
   adopting different prescriptions for the stellar lifetimes differ mostly in 
   the predicted evolution for those species originating mostly from low-mass 
   stars. We analyse the evolution of $^3$He and $^7$Li and show that it is 
   better reproduced if adopting long lifetimes for the low stellar mass range.

   Oxygen abundance data recently derived for F and G dwarf stars in the solar 
   neighbourhood indicate that the [O/Fe] trend at super-solar [Fe/H] 
   continues downward, which is in concordance with models of Galactic 
   chemical evolution unless stellar lifetimes from Tinsley (1980) are 
   adopted. Notice that her prescriptions are characterized by very short 
   lifetimes for massive stars. 

   We conclude that, given the uncertainties still associated to current 
   observations, the main conclusions reached by chemical evolution models for 
   the Galaxy are left unchanged. However, it is clear that only further 
   observations and studying galaxies of different morphological type will 
   allow us to draw a firmer picture and a better understanding of the problem.

\begin{acknowledgements}
   We thank H. Rocha-Pinto for interesting informations and the referee, N. 
   Arimoto, for useful suggestions. DR and MT acknowledge the illuminating 
   discussions with J. Geiss and the members of the ISSI-LoLaGE team. The work 
   of DR has been partially funded by the Italian Space Agency through grant 
   IR\,11301\,ZAM and by INAF. Financial support from MIUR/COFIN 2003029437 
   and MIUR/COFIN 2003028039 is also acknowledged.
\end{acknowledgements}

\end{document}